\documentclass[twocolumn,amsmath,amssymb,english,superscriptaddress]{revtex4-1}
\usepackage{graphicx,natbib,subfigure,tabularx,epsfig,longtable,amsfonts,rotating,subfigure,hyperref,babel,currfile}
\usepackage{color}

\newcommand{\bea}{\begin{eqnarray}}
\newcommand{\eea}{\end{eqnarray}}

\newcommand{\be}{\begin{equation}}
\newcommand{\ee}{\end{equation}}


\begin{document}
\title{The influence of localization transition on dynamical properties for an extended Aubry-Andr\'e-Harper model}
\author{X. L. Zhao}
\affiliation{School of Physics and Optoelectronic Technology, Dalian
	University of Technology, Dalian 116024, China\\}
\affiliation{Center for Quantum Sciences and School of Physics, Northeast Normal University, Changchun 130024, China\\}
\author{Z. C. Shi}
\affiliation{Department of Physics, Fuzhou University, Fuzhou 350002, China\\}
\affiliation{Fujian Key Laboratory of Quantum Information and Quantum Optics\\
	(Fuzhou University), Fuzhou 350116, China}
\author{C. S. Yu}
\affiliation{School of Physics and Optoelectronic Technology, Dalian
	University of Technology, Dalian 116024, China\\}
\author{X. X. Yi}\thanks{yixx@nenu.edu.cn}
\affiliation{Center for Quantum Sciences and School of Physics, Northeast Normal University, Changchun 130024, China\\}

\date{\today}
\begin{abstract}
We show the localization transition and its effect on two dynamical processes for an extended Aubry-Andr\'e-Harper
model with incommensurate on-site and hopping potentials. After specifying an extended Aubry-Andr\'e-Harper model,
we check the localization transition for all the eigenstates and  eigenenergy band splitting behavior versus a
system parameter.  To examine the effect of localization transition on dynamical processes, firstly, the slowly
pumping of the edge states are examined. In the dynamical processes, the system acts as conductor for the excitation
in the nonlocal region and insulator in the localized region. Then by quantum Lyapunov control method with different control Hamiltonians, we  prepare an edge localized state which exists in the nonlocal region. Compared to that 
in the nonlocal region, the control effect is suppressed in the localized region. Then we employ the entropy and occupation imbalance between even and odd sites to indicate the localization transition further. Finally, the 
experimental schemes based on cold atoms trapped quasiperiodic optical lattice and coupled optical waveguide arrays 
are suggested.
\end{abstract}	
\maketitle
\section{Introduction}
Localization is an intriguing phenomenon in many-body systems\cite{rmp57287}. For example, Aderson 
localization explains the transport mechanism for metal-insulator phase transition in solids~\cite{pr1091492}. 
This localization is interpreted as the interference effect of the electronic wave functions in presence of disorder.

Analogous but different to the Aderson localization in one dimension (1D) version~\cite{ppslsa68874,aips3133,prb142239,prl49833,njp670,Nature453895,prl511198,prb408225},
localization transition occurs in the Aubry-Andr\'e-Harper (AAH) model~ with on-site incommensurate  modulations\cite{ppslsa68874,aips3133}. This is a model intermediate between random and periodic systems with quasiperiodic modulated on-site potentials. It may reduce from the description of a two-dimensional (2D) quantum 
Hall system by using Landau gauge for the magnetic field~\cite{prb142239}.  The critical point for the localization transition is related to the ratio of the on-site and the hopping amplitudes. In a subsequent of variants of this 
model, extensive arguments about localization properties have been discussed
~\cite{prl501873,prl622714,Nature453895,Science349842,NP13460,prl103013901,prl109106402,prl109116404,prb5011365,prb284272,prb428282,prl752762,prl110076403}

AAH model has attracted wide attention since it can be used to study topological properties for the 2D counterparts~\cite{prl109106402,prl109116404,prl110076403}.
But the intensity of the magnetic field sets the obstacle for this model to be explored in solid systems 
currently. Fortunately, the developing quantum simulation systems\cite{RMP80885,NP8267} offer platforms to 
implement such models such as by non-interacting $^{39}$K Bose-Einstein condensate in quasirandom
optical lattice\cite{Nature453895} or interacting fermions in a one-dimensional quasirandom optical lattice\cite{Science349842,NP13460}. Besides, coupled optical waveguide arrays are also a candidate to 
implement such a model with tunable on-site and hopping parameters
\cite{prl103013901,prl109106402,oe146055,jpb43163001,prl102153901,prl100094101,nature424817,prl100013906}.

In this work, we consider an extended AAH model with the hopping and on-site potentials are both modulated incommensurately. We use the inverse participation ratio (IPR)to indicate the degree of localization. Average 
localization diagrams versus two modulation phases are exhibited. Then we specify a system to show the 
localization transition in terms of all the eigenstates and the energy spectrum. The splitting behavior of 
the energy spectrum coincides with the localization transition of the eigenstates. In addition, this transition 
may have significant effect on the dynamical properties of the system. Thus we show the influence of the 
localization on slowly pumping of an edge state; preparing the edge state by Lyapunov control methods. It is 
shown that this model acts as conductor for the excitation on  the chain in the nonlocal region, however as 
insulator in the localized region which blocks the transportation processes. However there may be various 
kinds of distribution patterns of the states on the chain which IPR may not reflect accurately. Thus we would 
use occupation imbalance between even and odd sites and the entropy to reveal the localization transition~\cite{Science349842,NP13460}. It can be seen that the localization transition behavior coincides 
with the one IPR indicates. Finally we propose the discussions may be checked in the system consist of cold  
atoms trapped in quasiperiodic optical lattice\cite{Nature453895,Science349842,NP13460} or coupled optical waveguide arrays\cite{prl103013901,prl109106402,oe146055,jpb43163001,prl102153901,prl100094101,nature424817,prl100013906}.

Since the Lyapunov control method would be employed to prepare the edge state, we give a brief introduction to 
this method here. It is a `close-loop design, open-loop apply' strategy which has been applied to finish control 
goals in various systems effectively~\cite{42IEEE,Auto411987,PRA80052316,PLA378699,sr513777}. In this method, after choosing suitable control Hamiltonians based on the controlled systems, the control fields are designed based on 
Lyapunov functions. Then the system is driven to the goal state. According to LaSalle' invariant principle~
\cite{Lasalle}, the control fields tend to vanish when the system is asymptotically steered to the target. In this 
work, we daopt one Lyapunov function with two different control  Hamiltonians to design the control fields to 
prepare an intriguing edge state.

This work is organized as follows. In section~\ref{AAHMODEL}, we show the average localization for all the 
eigenstates versus system parameters and specify the extended AAH model which will be discussed further. Then 
we show that the splitting behavior for the energy spectrum coincides with the localization transition of the 
eigenstates. Then in section~\ref{DYNAMICAL}, we consider the influence of the localization to adiabatic pumping 
of the edge states;  the procedure to prepare the edge localized state by state-distance Lyapunov control method
 with two different control Hamiltonians. Then in section \ref{DISCUSSIONS}, to further reveal the localization 
 patterns on the chain, besides IPR, we employ entropy of the eigenstates and occupation imbalance between even 
 and odd sites to show  the localization behavior. Then we provide the experimental schemes to realize this model 
 based on cold atoms in quasiperiodic optical lattice and coupled optical waveguide arrays. Finally, we summarize in section~\ref{SUM}.
\section{The extended AAH model}
\label{AAHMODEL}
In this work, we consider an extended 1D AAH model with both incommensurate on-site potentials and nearest-neighbor 
hopping interactions which is described by the following tight-binding Hamiltonian ($\hbar=1$ is assumed in this 
work):
\bea
\hat{H}_0&=&\sum_{n=1}^{N} v \cos(2\pi b n+\theta_{v})\hat{c}^{\dagger}_{n}\hat{c}_n+\nonumber\\
&&\sum_{n=1}^{N-1} [\lambda_0+\lambda \cos(2\pi b n+\theta)] \hat{c}^{\dagger}_{n+1}\hat{c}_n+h.c.,
\label{eq:Hamiltonian}
\eea
 where  $\hat{c}^\dagger_n$ and $\hat{c}_n$ are the polarized Fermionic creation and annihilation operators 
 for site $n$. $N$ is the total number of sites on the 1D AAH chain. The terms in the first line in (\ref{eq:Hamiltonian}) 
 represent the incommensurate on-site potentials where $v$ is the amplitude. The last terms describe the kinetic 
 energy from the nearest-neighbor hopping. $\lambda_0$ is the amplitude which is taken as the energy unit 
 throughout this  work and  $\lambda$ indicates the modulation amplitude. Even this Hamiltonian is a mutation 
 of the quasiperiodic model derived from the tight-binding square lattice with next-nearest couplings 
 \cite{prb5011365,prb284272,prb428282,prl752762}, it can be implemented by cold atoms trapped in quasiperiodic
  optical lattice\cite{Nature453895,Science349842,NP13460} or coupled optical waveguide arrays
  \cite{prl103013901,prl109106402,oe146055,jpb43163001,prl102153901,prl100094101,nature424817,prl100013906}.

Note that whether $b$ is rational or not influences the localization properties of such kinds of systems~\cite{prb5011365,prb284272,prb428282,prl752762}. When $b$ is a rational number, the adjustable on-site 
potentials have the periodicity of $1/b$ which is determined by the magnetic field penetrating the 2D partner 
of the 1D chain~\cite{ppslsa68874,prb142239}. Due to Bloch theorem, no localization transition occurs in this 
case. Whereas in this work, we will focus on the case that $b$ is the irrational number $(\sqrt{5}-1)/2$, 
namely, the inverse of golden mean. Mathematically, localization occurs with $b$ being an irrational Diophantine number\cite{AM1501159}. We will exhibit the localization transition and its influence on three dynamical processes 
in the following.
 \begin{center}
 	\begin{figure}[htb!]
 		\includegraphics[scale=0.36]{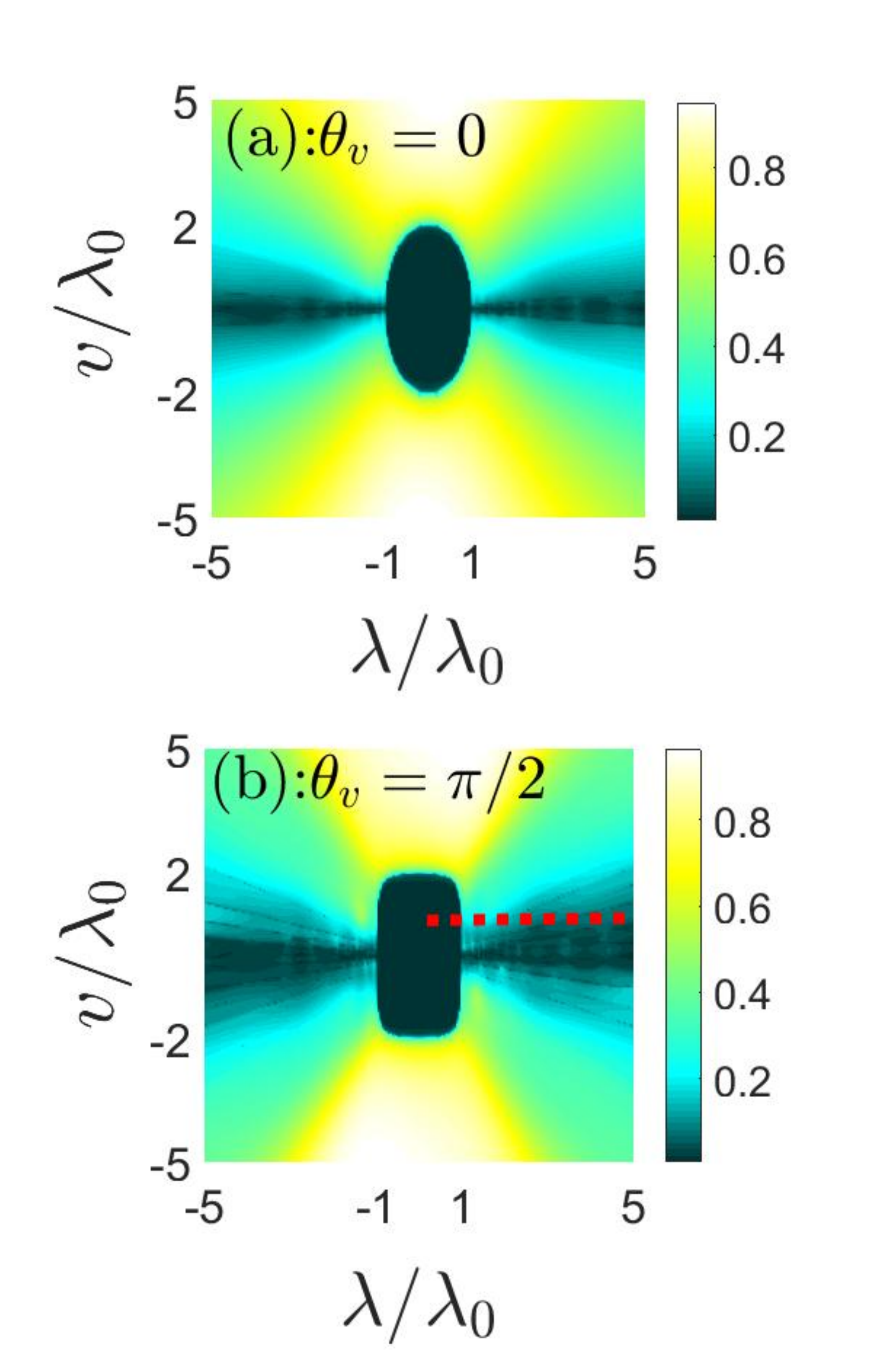}
 		\vspace{-0.5cm}
 		\caption{The average localization transition diagrams indicated by IPR over all the eigenstates for 
$\theta_v=0$ and $\theta_v=\pi/2$ when $\theta=0$. When $v$=0, only hopping potential exists which leads to 
nonlocal region. We have used a chain composed of 40 sites. The other parameter are $b$=$(\sqrt{5}-1)/2$ and  $v/\lambda_0$=1 in the Hamiltonian~(\ref{eq:Hamiltonian}). We would check the localization transition for a 
range of $\lambda/\lambda_0$ when $v/\lambda_0=1$ as the red dotted line shows in (b).}
 		\label{f:AdersonphaseT}
 	\end{figure}
 \end{center}
 \subsection{Localization transition and fractal energy band}
We review the localization transition to draw forth the system we will discuss further. To indicate the degree 
of localization for a state, the inverse participation ratio IPR=$\sum_n|\psi(n)|^4$ is employed~\cite{pr1393}, 
where $\psi(n)$ are the amplitudes for a state on site $n$. It can be seen that when $\psi(n)$ distributes 
homogeneously over $N$ sites on a chain, namely $|\psi(n)|\sim1/\sqrt{N}$, then IPR $\sim1/N$. Whereas, if $\psi(n)$
locates mainly over a range $\zeta$, namely $|\psi(n)|\sim1/\sqrt{\zeta}$, then IPR $\sim1/\zeta$. Thus IPR tends 
to vanishing for nonlocal states but finite values for localized states while $N$ is large. Considering there may 
be various kinds of localization patterns on the chain, to reveal the localization further, we exhibit that both 
entropy for the eigenstates and occupation imbalance between the even and odd sites\cite{Science349842,NP13460} 
confirm the localization transition IPR reflects.

First we check the average localization over all the eigenstates as a function of $\lambda/\lambda_0$ and 
$v/\lambda_0$ in Hamiltonian (\ref{eq:Hamiltonian}) for $\theta_v=0$ and $\pi/2$ in Fig.~\ref{f:AdersonphaseT}. 
It can be seen that when $v$=0, only hopping interactions remain which leads to the eigenstate being nonlocal for
$\lambda$s. For finite $v/\lambda_0$,  with increasing of $\lambda/\lambda_0$, the nonlocal region tends to expand 
as the shaded area shows in Fig.~\ref{f:AdersonphaseT}. This may result from that the hopping leads to diffusion 
of the distributions. When $\theta_v=0$, the boundary for localization transition as shown in Fig.\ref{f:AdersonphaseT}(a)  fulfills the condition $(v/2)^2+\lambda^2=\lambda_0^2$.  However when 
$\theta_v=\pi/2$, the boundary tends to be a rectangle as in Fig.~\ref{f:AdersonphaseT}(b). In this work we will 
focus on the extended AAH model with $v/\lambda_0=1$.
\begin{center}
	\begin{figure}
		\centering
		{\includegraphics[scale=0.45]{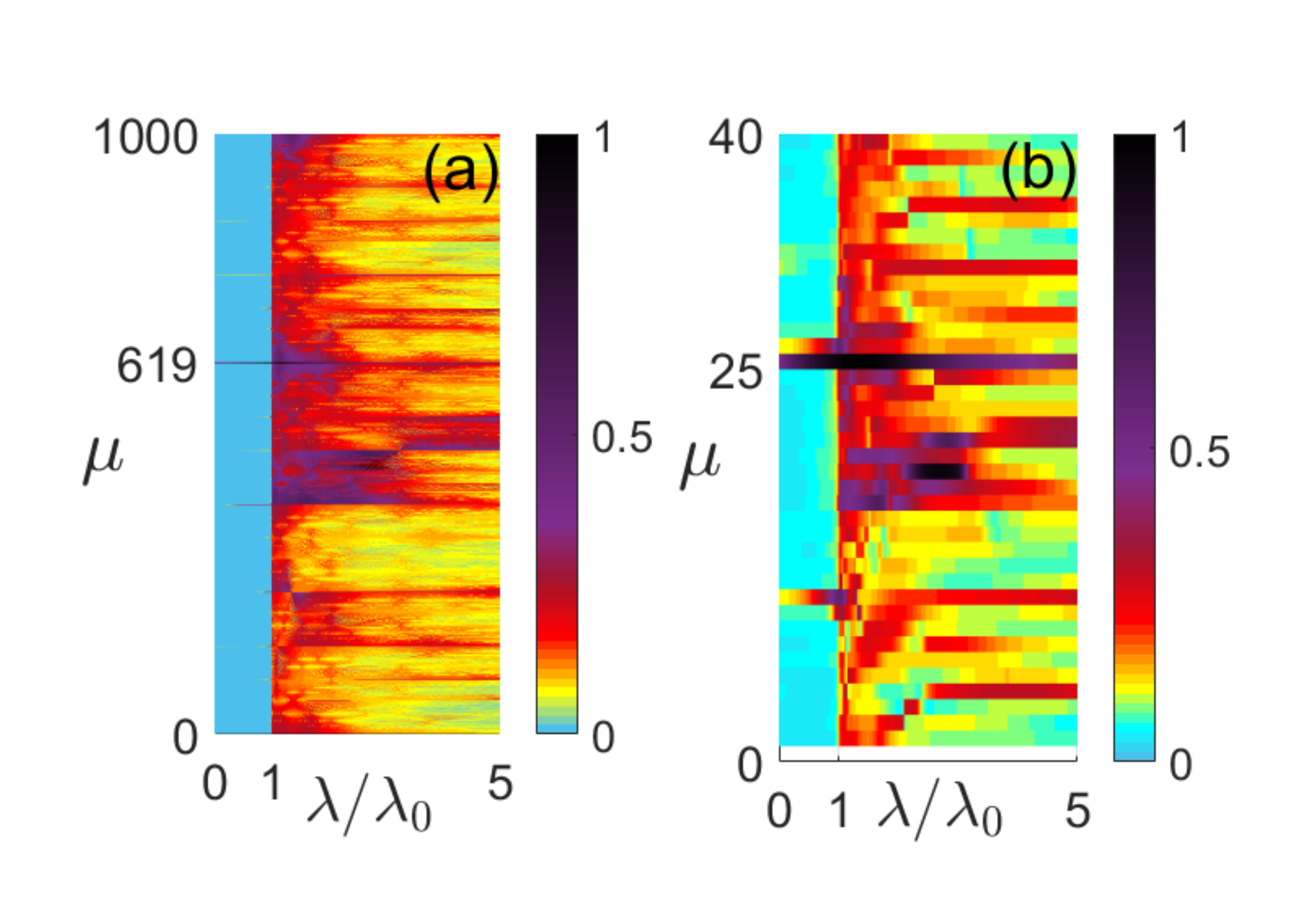}}
		\vspace{-0.5cm}
		\caption{The IPR for the eigenstates of a chain for a range of $\lambda/\lambda_0$ when $b=(\sqrt{5}-1)/2\approxeq0.618$. We show (a): $N$=1000  and (b): $N$=40  sites on the 
			chain for comparing when $\theta=0$, $\theta_v=\pi/2$ and $v/\lambda_0=1$. With increasing of 		$\lambda/\lambda_0$, the IPR for some eigenstates decrease due to the hopping interaction dominates.  
			The eigenstates are denoted by index $\mu$.}
		\label{f:AdersonIPR}
	\end{figure}
\end{center}

\begin{center}
	\begin{figure}
		\centering
		{\includegraphics[scale=0.29]{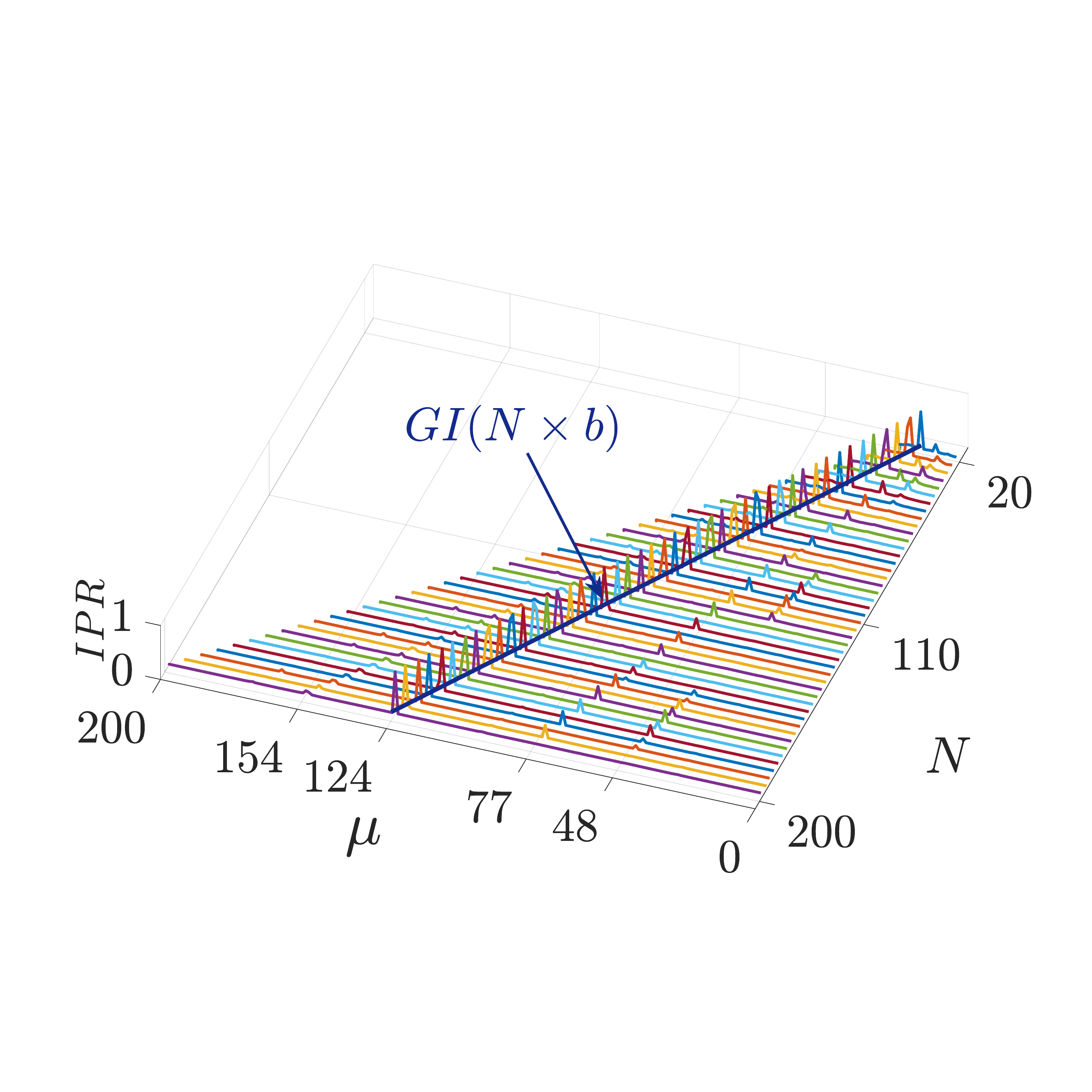}}
		\vspace{-1cm}
		\caption{The IPR for the eigenstates (denoted by $\mu$) in a range of length (denoted by $N$) in the 
			nonlocal region with $\lambda/\lambda_0=0.5$. It can be seen that the most obvious localized at
			GI$(N\times b)$, for example, GI$(200\times b)$=124. The position of the other localized  eigenstates
			are related to inverse of the golden mean $b=(\sqrt{5}-1)/2$.  The other parameters are $v/\lambda_0=1$, $\theta=0$ and $\theta_v=\pi/2$.}
		\label{f:AdersonLV}
	\end{figure}
\end{center}

Then we show the localization transition diagram for all the eigenstates versus $\lambda/\lambda_0$ by two examples 
with different lengths in Fig.\ref{f:AdersonIPR} when $\theta_v=\pi/2$, $v/\lambda_0=1$ and $\theta=0$. It can be 
seen that the localization transition occurs not only in the ground state but also in the excited states which
 hints that this transition can occur in relatively high temperature. According to Fig.~\ref{f:AdersonphaseT}(b),  $|\lambda/\lambda_0|\simeq1$ is the transition point for localization when $|v/\lambda_0|\lesssim2$. The transition 
 of localization is clear and yet there are localized states in the nonlocal region. In particular the location for 
 the most obvious localized state (the states are ordered  according to the corresponding eigenenergies from small 
 to large) is related to the inverse of golden mean $b$=$(\sqrt{5}-1)/2$, independent on the length of  the AAH 
 chain. We find that the location of the most obvious localized state in the nonlocal region can be expressed as GI$(N\times b)$. Here GI$(x)$ denotes the function which returns the minimum integer among those larger 
 than $x$ and $N$ is the number of sites on the chain.  For example, for the chain with 1000 sites: 
 GI$(1000\times b)$=619 and 40 sites: GI$(40\times b)$=25, and so on. Yet there are other eigenstates related to 
 $b$ in the Hamiltonian which has similar localization character but not obviously localized in the nonlocal region. 
 We show the IPR for the eigenstates in the nonlocal region ($\lambda/\lambda_0=0.5$) of different lengths in Fig.\ref{f:AdersonLV}. It can be seen that the GI$(N\times b)$th states locate obviously in the nonlocal region for 
 the range of lengths. However, explanations for these elusive localized states may need further work.
\begin{center}
	\begin{figure}
		\centering
		{\includegraphics[scale=0.2]{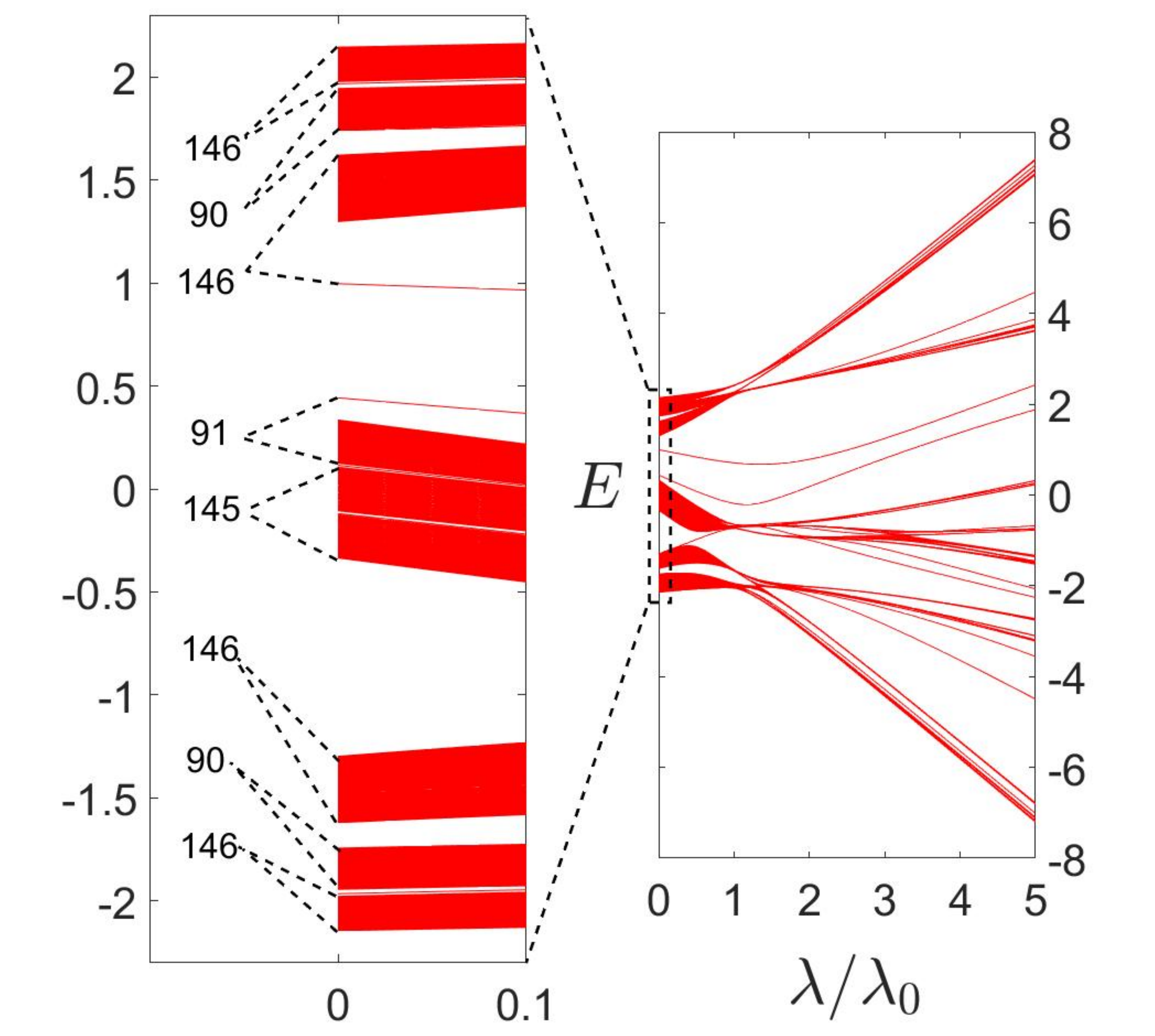}}
		\vspace{0cm}
		\caption{The energy spectrum as a function of $\lambda/\lambda_0$ when $v/\lambda_0=1$, $\theta_v=\pi/2$ 
			and $\theta$=0. We have set $N$=1000 for the chain. The energy band is zoom in to show that the energy 
			is splitting in a  way of fractal related to the inverse of the golden mean $(\sqrt{5}-1)/2$. The numbers denote the amount of energy levels in the bands. The change of the splitting behavior coincides with the localization transition in	Fig.\ref{f:AdersonIPR} (a).}
		\label{f:Adersonband}
	\end{figure}
\end{center}

\begin{center}
	\begin{figure}
		\centering
		{\includegraphics[scale=0.18]{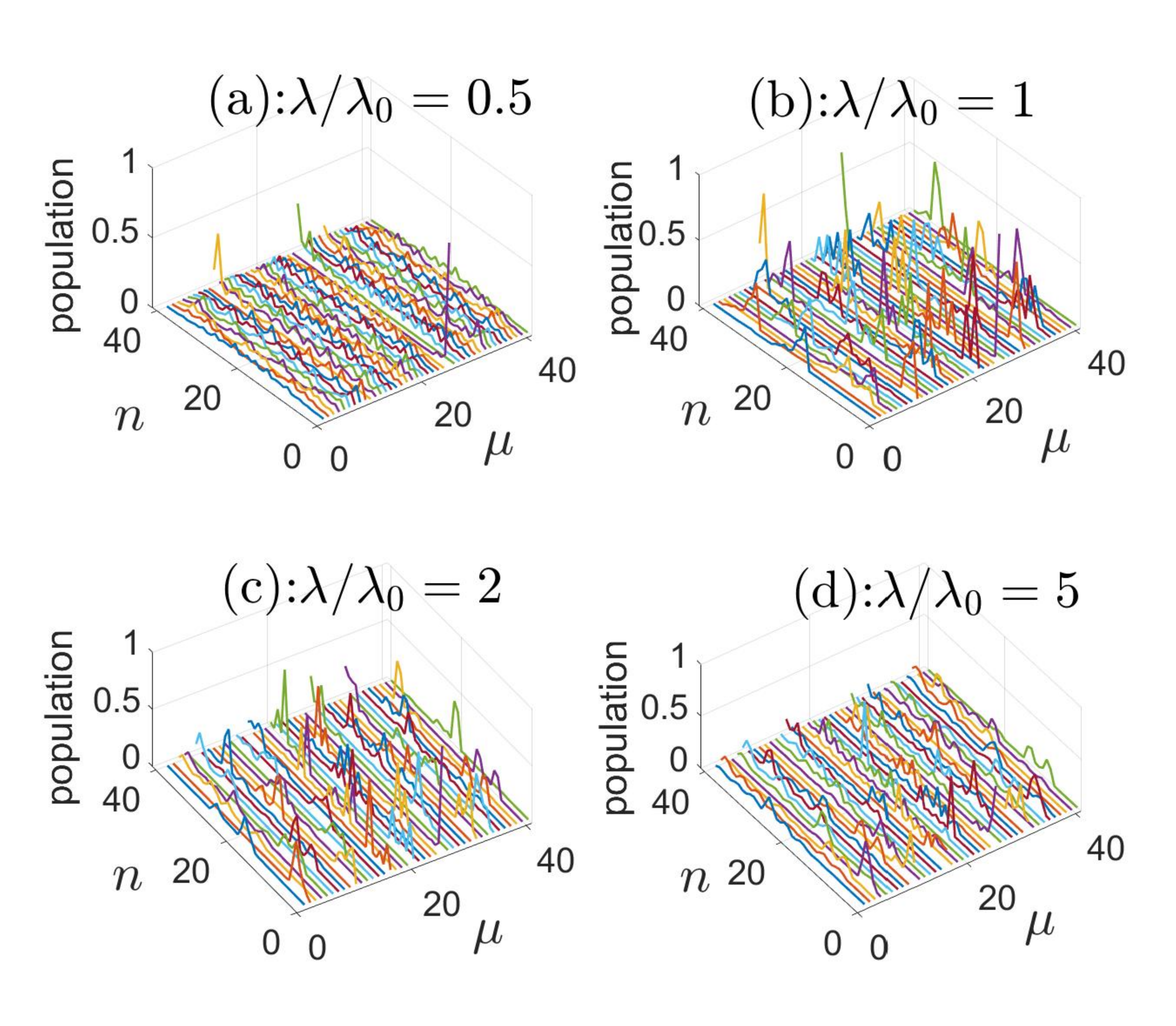}}
		\vspace{-0.5cm}
		\caption{The population of the eigenvectors for $\lambda/\lambda_0$=0.5, 1, 2 and 5 for a chain with $N=40$ 
			sites. The degree of localization for $\lambda/\lambda_0=5$ seems less than those of $\lambda/\lambda_0=$2 
			and 1 which coincides with that as Fig.~\ref{f:AdersonphaseT} and Fig.~\ref{f:AdersonIPR} show. This is because with increasing of $\lambda/\lambda_0$ , the hopping dominates. $n$ and $\mu$ denote the position
			 of the sites on the chain and index for the eigenstates respectively. The other parameters are same to those in Fig.~\ref{f:AdersonIPR}(b).}
		\label{f:Adersonpop}
	\end{figure}
\end{center}

The localization properties and self-similar structures of distributions for eigenstates at the localization boundary 
in similar systems has been investigated widely~\cite{prb334936,jpcssp205999,epl4597}. In this work, we find that the energy band versus $\lambda/\lambda_0$ splits in a fractal way in the nonlocal region. And the splitting behavior in 
the nonlocal region is different to that in the local region. Then we check the energy band versus $\lambda/\lambda_0$
 in Fig.~\ref{f:Adersonband}. It can be seen that the energy band splits in the nonlocal region with a fractal structure related to the inverse of golden mean. For example,  for the chain of $N=$1000 as in Fig.~\ref{f:Adersonband}, a simple examine is: 90/146$\simeq$(146)/(90+146)$\simeq$(90+146)/(146+90+146)...$\simeq$ 0.618 which approximates $b$=$(\sqrt{5}-1)/2$. With increasing of $N$, the ratios would approach to $(\sqrt{5}-1)/2$ which is elusive. But the energy levels in the bands tend to be bunching and crossing when $\lambda/\lambda_0$ strides over 1. The bunching 
 and crossing of the energy levels correspond to the localization of the eigenstates which can be confirmed in Fig.\ref{f:Adersonpop}. This change of the splitting behavior coincides with the localization transition in Fig.~\ref{f:AdersonIPR}. 

To show the localization in detail,  the distributions for different eigenstates are plotted in Fig.~\ref{f:Adersonpop}
 by examples. It can be seen that in the nonlocal region, most of the eigenstates are extended except for 
 the obvious edge-localized GI$(N\times b)$th state.  When $\lambda/\lambda_0$ is larger than 1, all the eigenstates 
 are localized. Later, one can see that the localization transition influences dynamical processes based on the 
 transport properties of the excitations on the chain. The dynamical processes are adiabatic pumping for an edge 
 state and  preparing the edge state by Lyapunov control method.
\section{Influence of the localization on the dynamical processes}\label{DYNAMICAL}
\subsection{Aadiabatic pumping for the edge localized state}
\begin{center}
	\begin{figure}
		\centering
		{\includegraphics[scale=0.15]{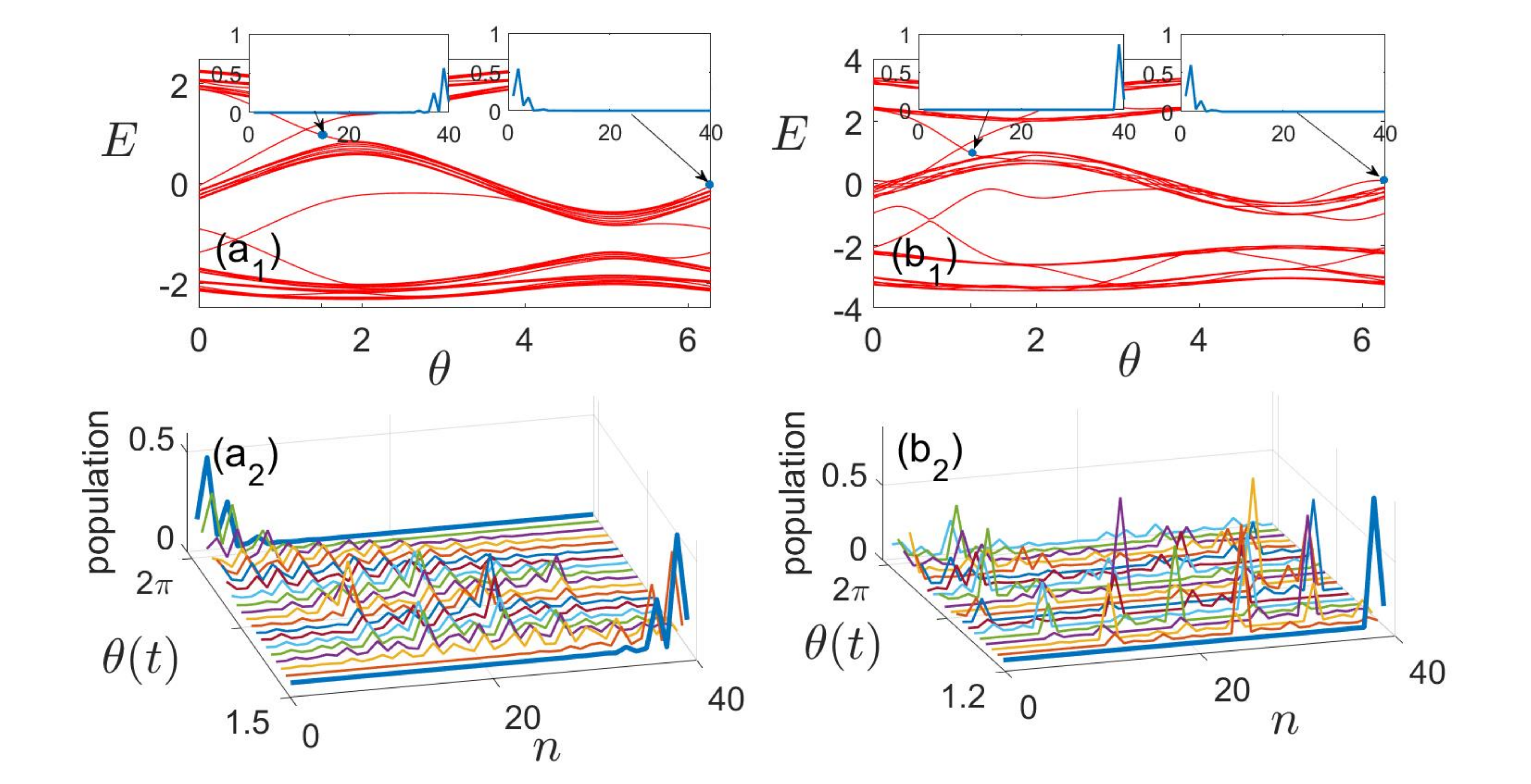}}
		\vspace{-0.5cm}
		\caption{($a_1$): The energy spectrum as a function of $\theta$ for $\lambda/\lambda_0$=0.8 where the
			 phase $\theta_v=0$. ($a_2$) : the pumping of the 25th eigenstate by slowly scan of $\theta$(t)= 1.5+t(2$\pi$-1.5)/$\tau$.  ($b_1$) and ($b_2$) are those when $\lambda/\lambda_0=2$ 
			 for the slowly varying $\theta$(t)=1.2+t(2$\pi$-1.2)/$\tau$.  $t$ denotes evolving time in units of $1/\lambda_0$ and $\tau$ denotes the rate for the pumping processes. The larger $\tau$ is, the more 
			 slowly of the pumping. The inserts in $a_1$ and $b_1$ are the distributions for the eigenstates as 
			 the arrows point to. We have set $\tau=10^5$ in these pumping processes. The other system parameters
			  are same to those in Fig.\ref{f:AdersonIPR}(b).}
		\label{f:Adersontheta}
	\end{figure}
\end{center}

\begin{center}
	\begin{figure}
		\centering
		{\includegraphics[scale=0.15]{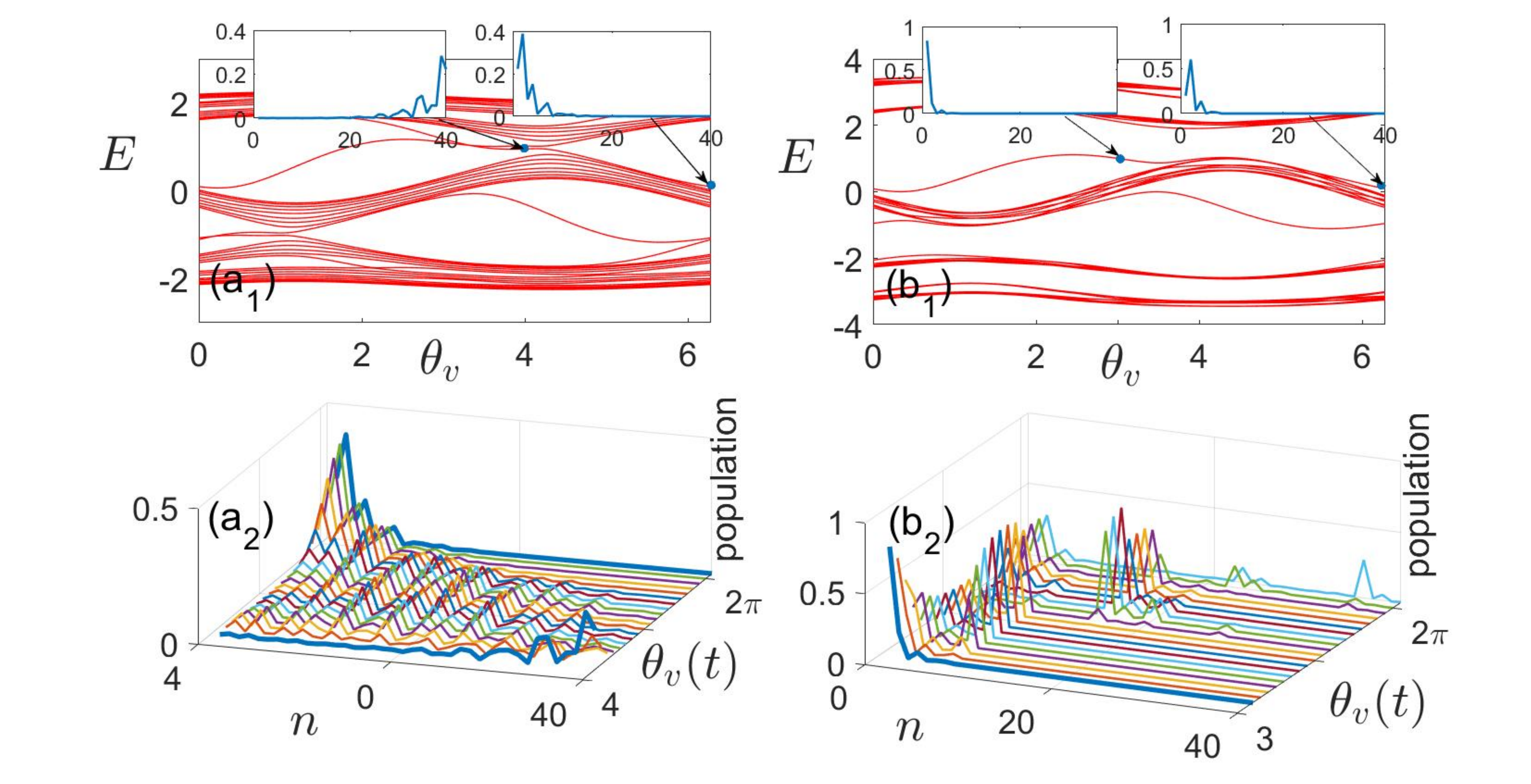}}
		\vspace{-0.5cm}
		\caption{($a_1$): The energy spectrum as a function of $\theta_v$ for $\lambda$=0.5 where the phase 
			$\theta=0$. ($a_2$) : the pumping of the 25th eigenstate by slowly scan of $\theta_v$(t)= 4+t(2$\pi$-4)/$\tau$. ($b_1$) and ($b_2$) are those when $\lambda=2$ for the slowly varying
			$\theta_v$(t)= 3+t(2$\pi$-3)/$\tau$. $t$ and $\tau$ have the same meaning in Fig.\ref{f:Adersontheta}. 
			The inserts in $a_1$ and $b_1$ are the distributions for the eigenstates as the arrows point to.
			 We have set $\tau=10^5$. The other system parameters are same to those in Fig.\ref{f:AdersonIPR}(b).}
		\label{f:Adersonphi}
	\end{figure}
\end{center}

The appearance of edge localized states whose eigenenergies span the band gap connecting different energy bands is 
usually treated as a signature for nontrivial topological character. Topological phase with edge states has been 
discussed in quasicrystals which is attributed to higher dimensional systems~\cite{prl109106402}. The AAH model in 
this work is an extension for such a quasicrystal chain which can be implemented by cold atom system trapped in quasiperiodic potentials\cite{Nature453895,Science349842,NP13460} or coupled optical waveguide array system\cite{prl103013901,prl109106402}. When $\theta$ or $\theta_v$ is regarded as an additional compact dimension,
some topological properties can be checked by adiabatic pumping of the edge states. Since there is localization 
transition in this system, we check the influence of this transition on the pumping processes for the edge states 
by slowly varying $\theta$ and $\theta_v$ in Fig.~\ref{f:Adersontheta} and Fig.~\ref{f:Adersonphi} respectively. It 
may approximate to an adiabatic process  when the pumping is slow enough.  Such pumping processes can be checked by solving the Schr\"odinger equations :
\begin{eqnarray}
\label{eq:linear}
i\partial_t\psi_n&=& [\lambda_0+\lambda \cos(2\pi b(n-1)+\theta(t))]\psi_{n-1}\\ \nonumber
&&+ [\lambda_0+\lambda \cos(2\pi b(n)+\theta(t))]\psi_{n+1}\\ \nonumber
&&  +v \cos(2\pi b n+\theta_{v}(t))\psi_n,
\end{eqnarray}
where $\psi_{n}$ ($n=1,2,...N$) are the amplitudes for the wavefunction on site $n$. To approximate the adiabatic 
pumping process, we assume that:  $\theta(t)= 1.2+t(2\pi-1.2)/\tau$ ($\theta_v$ fixed) or $\theta_v(t)= 3+t(2\pi-3)/\tau$ ($\theta$ fixed). Here $t$ denotes the evolution time. The time is in units of $1/\lambda_0$ here and after. $\tau$ determines the speed of evolution. The larger $\tau$ is, the more slowly the evolution will be, namely, the more degree of the process verges on an adiabatic one.
\begin{center}
	\begin{figure}
		\centering
		{\includegraphics[scale=0.15]{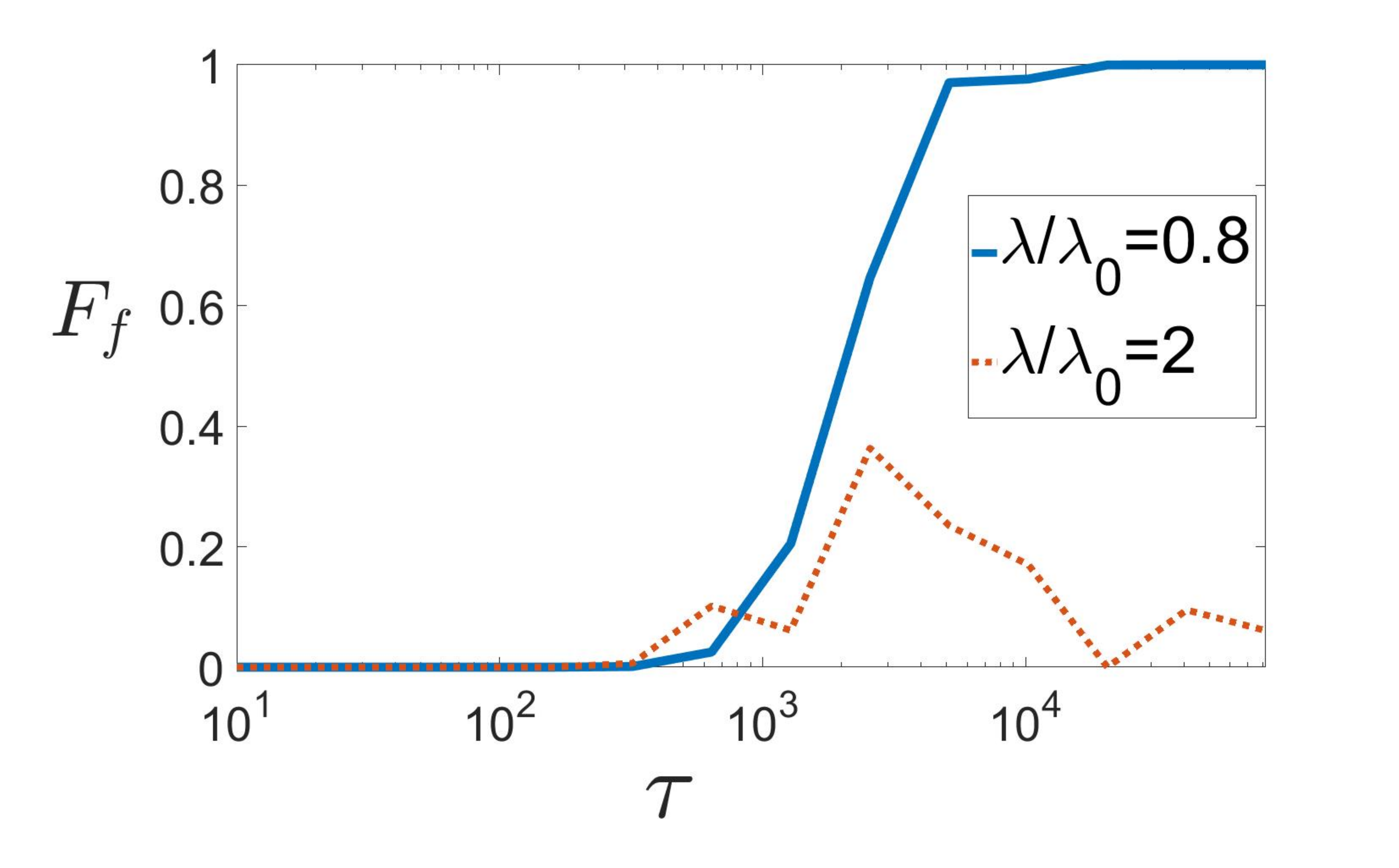}}
		\vspace{-1cm}
		\caption{The degree of adiabatic pumping as a function of $\tau$ is reflected by the fidelity between 
			the evolved state and the goal state in the nonlocal region ($\lambda/\lambda_0=0.8$) and localized 
			region ($\lambda/\lambda_0=2$) corresponding to the pumping processes in Fig.~\ref{f:Adersontheta}($a_2$)
			and ($b_2$) respectively. $F_f$ approaches 1 means the state is adiabatically evolved to the goal state successfully. The other parameters are same to those in Fig.\ref{f:Adersontheta}.}
		\label{f:Adersonadiabatic}
	\end{figure}
\end{center}

In terms of slowly varying $\theta$ or $\theta_v$, the energy spectrum are shown in ($a_1$)s and ($b_1$)s in Fig.~\ref{f:Adersontheta} and Fig.~\ref{f:Adersonphi} respectively. The crossing of the eigenenergy levels which 
connect two bands is a signature of topological phase transition. In terms of $\theta$, both in the extended 
and localized region, the edge states are topologically protected. For $\theta_v$, the edge states in the nonlocal 
region are topological protected. However in the localized region, the edge states which are pined to one end of 
the chain without topological phase transition.

 ($a_2$)s and ($b_2$)s in Fig.~\ref{f:Adersontheta} and Fig.~\ref{f:Adersonphi} show the slowly pumping for 
 the edge states in the nonlocal and local regions when $\theta(t)$ or $\theta_v(t)$ varies slowly. The influence 
 of localization on the pumping processes can be seen clearly. In the nonlocal region, as $\theta(t)$ or 
 $\theta_v(t)$ varies slowly, the topological protected edge state can be adiabatically pumped into a bulk band 
 and then become an edge state localized at the opposite end of the chain. However in the localized region, 
 the bulk states become localized and the adiabatic pumping processes are suppressed. The relation between 
 localization transition and topological phase transition may be investigated in future.

 According to quantum adiabatic theorem~\cite{pa51165,jpsj5435},  it is necessary that the system evolves slowly 
 enough to remain the state  as the instantaneous eigenstate of the time dependent Hamiltonian  and the gap between 
 the eigenvalues also play a critical  role. Whether the state is pumped to the goal state $|\varphi_{g}\rangle$
 successfully can be indicated by defining the fidelity as $F_f=|\langle\varphi(\tau)|\varphi_{g}\rangle|^2$, 
 where  $|\varphi(\tau)\rangle$ is the state when the pumping process finishes. We check the fidelity $F_f$ as 
 a function of $\tau$  in the nonlocal and localized region in Fig.\ref{f:Adersonadiabatic} which correspond 
 pumping processes in Fig.\ref{f:Adersontheta} ($a_2$) and ($b_2$). It confirms that $\theta(t)$ evolves slowly 
 enough is necessary for the state  being adiabatically pumped to the goal state successfully. And in the localized region, the failure of the pumping process may result from that the gaps between the eigenstate and the neighbors 
 in the local region are to narrow as shown in Fig.\ref{f:Adersonband} and energy levels cross in the localized 
 region. This follows the statement of adiabatic theorem~\cite{pa51165,jpsj5435}.

\begin{center}
	\begin{figure}
		\centering
		{\includegraphics[scale=0.24]{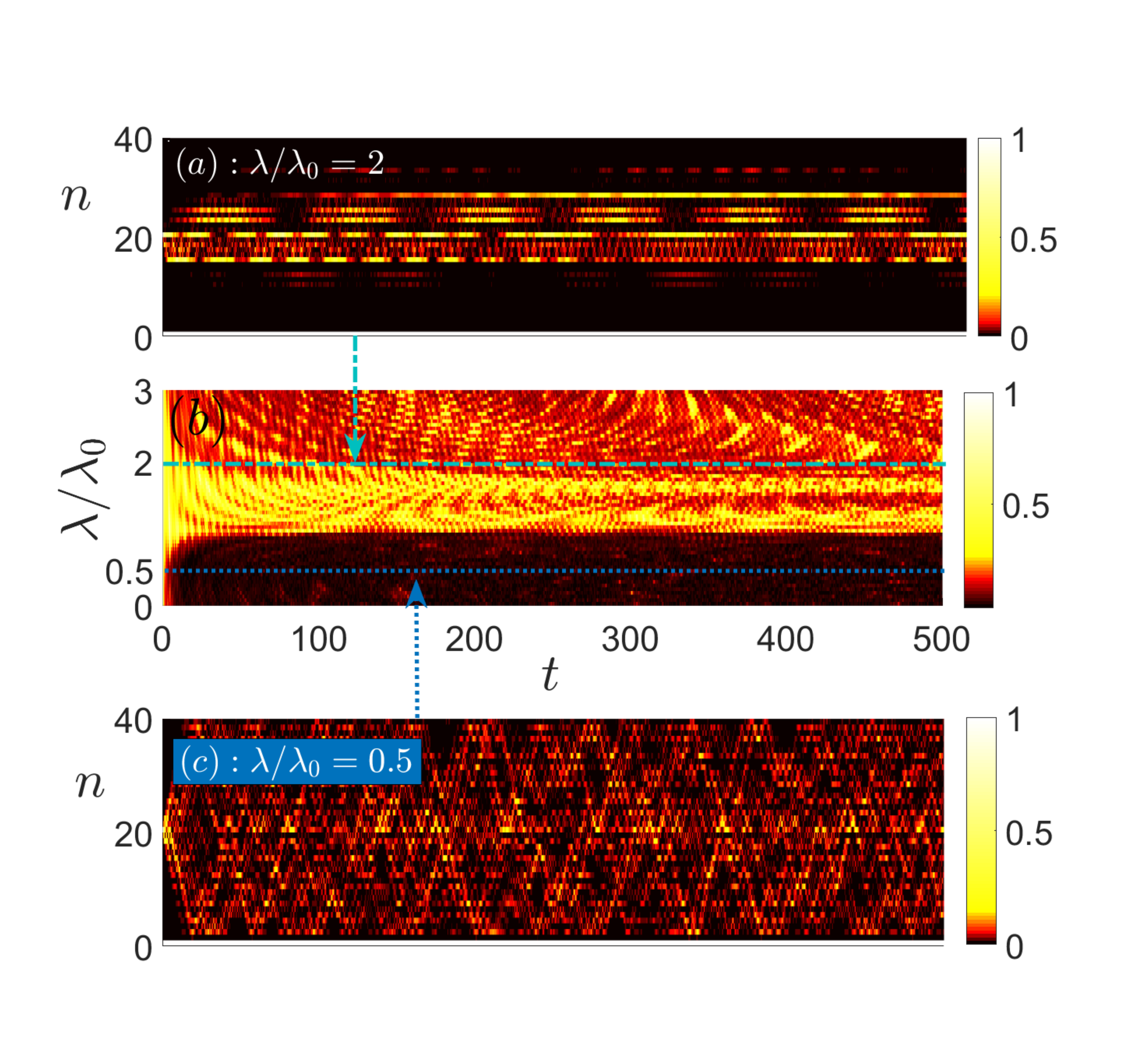}}
		\vspace{-1cm}
		\caption{ (b): The evolution of IPR for an excitation when the excitation initially locates at the middle 
			site on the chain versus $\lambda/\lambda_0$. (a): The evolution for the excitation when 
			$\lambda/\lambda_0$=2 and (c): the evolution for the excitation when $\lambda/\lambda_0$=0.5. The other parameters are $N=40$, $\theta=0$ and $\theta_v=\pi/2$.}
		\label{f:Adersondynamics}
	\end{figure}
\end{center}

The pumping processes are based on the transportation of the excitations on the chain. Thus in Fig.~\ref{f:Adersondynamics}, we show the transportation for the excitations on the chain in the two different localization regions respectively. It shows that in the nonlocal region, the excitations spread on the chain as it propagates. But in the localized region, the spreading is suppressed and obvious localized oscillation appears. From 
this point of view, one may concludes that the system acts as conductor in the nonlocal region and insulator in the localized region for the excitations. This may be the origin for the failure of adiabatic pumping for the edge states 
in the localized region. This also signifies that the  localization transition can be reflected by dynamical process 
of the excitations. Next we would examine the suppression of the control effect in the localization region compare 
to that in the nonlocal region when we use Lyapunov control method to prepare the edge state.
\subsection{Prepare  edge localized state by Lyapunov control}
\label{LyapunovC}

\begin{center}
	\begin{figure}[htb!]
		\centering
		\subfigure{\includegraphics[scale=0.2]{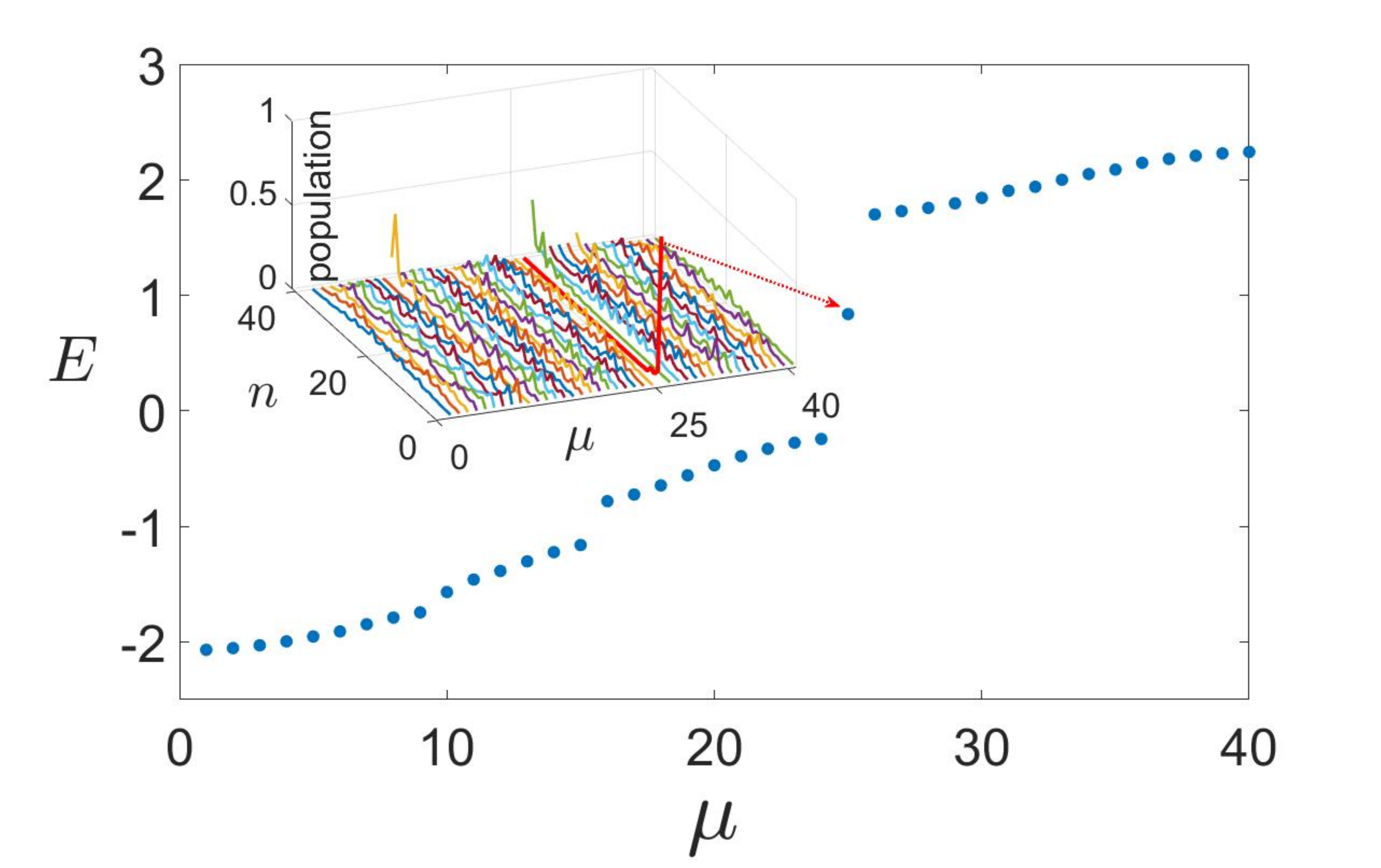}}
		\vspace{-0.5cm}
		\caption{The eigenenergies for the chain when $\theta=0$ and $\theta_v=\pi/2$ for the chain with $N$=40 
			when $\lambda/\lambda_0$=0.5. In this case, the 25th eigenstate is the edge localized state even in the nonlocal region with gap-located eigenenergy. We choose it as the control target to prepare by Lyapunov control method.}
		\label{f:Adersoneig}
	\end{figure}
\end{center}

One may want to positively influence the dynamics of a quantum system to achieve an object by control methods. 
Usually a specified state can be chosen as the control goal. In this extended AAH model, the localized edge state (GI$(N\times b)$th state) in the nonlocal region is an intriguing target which is shown in Fig.~\ref{f:Adersoneig}. 
Its corresponding eigenenergy locates in the gap apart from others. Quantum Lyapunov control method may be an 
effective tool to prepare such states since this method has been employed to prepare eigenstates effectively for 
many quantum systems~\cite{42IEEE,Auto411987,PRA80052316,PLA378699,sr513777}. Considering this control method is 
also based on transport properties of the excitations, we would check the influence of localization on the control 
effect.

In quantum Lyapunov control, in order to achieve a goal, control fields need to be designed based on Schr\"odinger equation: $i \partial_{t}|\psi(t)\rangle=(\hat{H}_0+\sum_nf_{Ln}(t)\cdot\hat{H}_{Cn})|\psi(t)\rangle$,
 where $\hat{H}_0$ and $\hat{H}_{Cn}$ denote the free Hamiltonian of the controlled system and the control 
 Hamiltonians, respectively. $f_{Ln}(t)$ are the control fields need to be designed. Generally, there are three 
 kinds of design schemes in this method~\cite{42IEEE,Auto411987,PRA80052316,PLA378699,sr513777}. Here we consider 
 the Hilbert-Schmidt state-distance scheme to prepare the edge localized eigenstate as shown in Fig.\ref{f:Adersoneig}.
  We denote  the target state as  $|\varphi_f\rangle$ here and the  instantaneous state at time $t$ during the control process as $|\varphi(t)\rangle$. According to the Hilbert-Schmidt state-distance scheme  in quantum Lyapunov control method, the Lyapunov function can be written as bellow:
\begin{eqnarray}
	V_{L}=\frac{1}{2}(1-|\langle \varphi_f|\varphi(t)\rangle|^{2}),
	\label{V1}
\end{eqnarray}
where $|\langle \varphi_f|\varphi(t)\rangle|^{2}$ denotes the transition probability from $|\varphi(t)\rangle$ to $|\varphi_f\rangle$. It reflects the distance between the states $|\varphi(t)\rangle$ and $|\varphi_f\rangle$ in 
the Hilbert space. According to the design process in Lyapunov control  method~\cite{42IEEE,Auto411987,PRA80052316,PLA378699,sr513777}, the first-order time derivative for $V_{L}$ need 
to be calculated as:
\begin{center}
	\begin{figure*}
		\centering
		\subfigure{\includegraphics[width=1.05\textwidth]{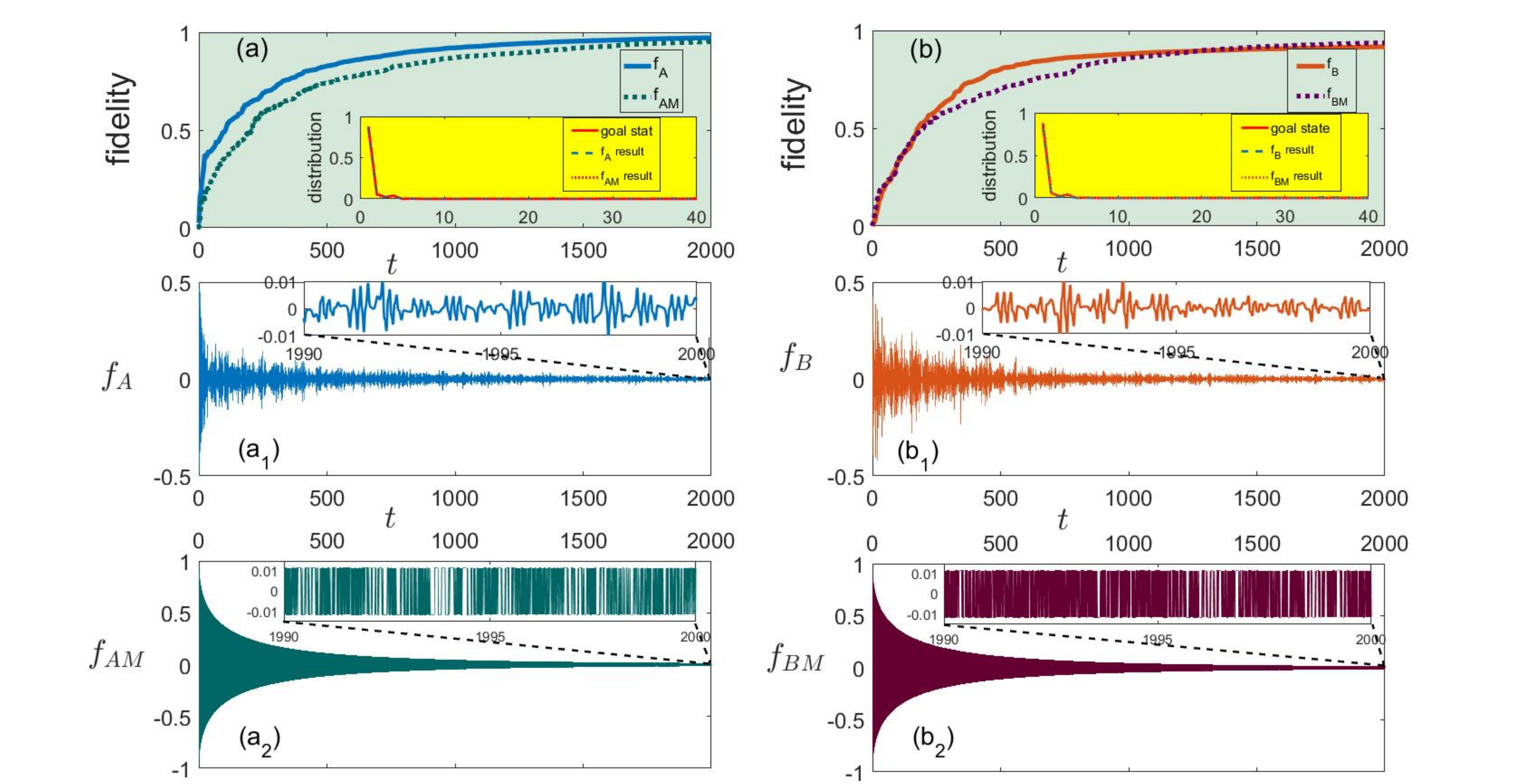}}
		\vspace{0cm}
		\caption{The fidelity versus time in the two Lyapunov control processes with hopping (Lyapunov control-A) 
			and on-site (Lyapunov control-B) control Hamiltonians. The insert in (a) and (b) show that the control 
			results match the goal well.  $f_A$ and $f_B$ exhibit the evolution for the designed control fields corresponding to the control Hamiltonians (\ref{hoppingHc}) and (\ref{onsiteHc}), respectively. $f_{AM}$
			 and $f_{BM}$ are the optimized control fields where we have set $\xi=0.1$ in (\ref{modulation}). We 
			 have chosen $V_c$=10, K=5, X=10 and Z=1 in the control Hamiltonians (\ref{hoppingHc}) and (\ref{onsiteHc}).
			The initial state is the one with all population on the 40th site. The other parameters are same to 
			those in Fig.~\ref{f:Adersoneig}.}
		\label{f:Adersonfid}
	\end{figure*}
\end{center}
\begin{eqnarray}
	\begin{aligned}
		\partial_t{V}_L=&-\sum_{n}f_{Ln}(t)\cdot|\langle\varphi(t)|\varphi_f\rangle|\times\\
		& Im[e^{i\arg\langle\varphi(t)|\varphi_f\rangle}\langle\varphi_f|\hat{H}_{Cn}|\varphi(t)\rangle],
		\label{Vdot}
	\end{aligned}
\end{eqnarray}
where we employ Im$[\bullet]$ to denote the imaginary part of $\bullet$. Next $f_{Ln}(t)$ need to be designed 
in order to achieve a control goal. According to the physical meaning of $V_L$ in~(\ref{V1}), the control fields $f_{Ln}(t)$  act in a way to make the distance between $|\varphi(t)\rangle$ and $|\varphi_f\rangle$ shrink. Thus a
succinct and valid choice is
\begin{eqnarray}
	\begin{aligned}
		f_{Ln}(t)=T_{n}\cdot Im[e^{i\arg\langle\varphi(t)|\varphi_f\rangle}\langle\varphi_f|\hat{H}_{Cn}|\varphi(t)\rangle],
	\end{aligned}\label{fd}
\end{eqnarray}
where $T_{n}>0$ are the free constants. When $\langle\varphi(t)|\varphi_f\rangle=0$, the angle $\arg\langle\varphi(t)|\varphi_f\rangle$ is uncertain. Without loss of generality,
we set $\arg\langle\varphi(t)|\varphi_f\rangle=0$ in this work.

Until now we have not specified the exact forms for the control Hamiltonians $\hat{H}_{Cn}$. Their forms are 
based on the controlled systems. For such an AAH chain, there may be various control Hamiltonians which can be
adopted to achieve the goal. Here we consider time dependent hopping-control Hamiltonian (Lyapunov control-A)
 withthe elements:
\begin{eqnarray}
	\begin{aligned}
		&\hat{H}_{ch}(m,n)\\
		&=\delta_{m,n+1}V_{c}\cos[2\pi K b m+X\cos(2\pi Z b \omega_{c}t)]\\
		&+h.c.,
		\label{hoppingHc}
	\end{aligned}
\end{eqnarray}
and  the time dependent onsite-control Hamiltonian (Lyapunov control-B) with the elements:
\begin{eqnarray}
	\begin{aligned}
		&\hat{H}_{co}(m,n)\\
		&=\delta_{m,n}V_{c}\cos[2\pi K b m+X\cos(2\pi Z b \omega_{c}t)],
		\label{onsiteHc}
	\end{aligned}
\end{eqnarray}
where $V_c$, K, X and Z are free control Hamiltonian parameters. The control Hamiltonians are all Hermitian 
to insure the evolution unitary and control fields real. Such kinds of control fields can be realized by  
electro-optic region modulator\cite{ PRL95170404,NATURE9769} while the model is implemented by cold atoms 
trapped in quasiperiodic optical lattice systems\cite{Nature453895,Science349842,NP13460}. When this model 
is implemented in coupled optical waveguide systems, the parameters can be tuned by changing on-site and hopping 
potentials along the propagation of the excitations along the waveguides~\cite{oe146055,jpb43163001,prl102153901}.
\begin{center}
	\begin{figure}[htb!]
		\centering
		\subfigure{\includegraphics[scale=0.32]{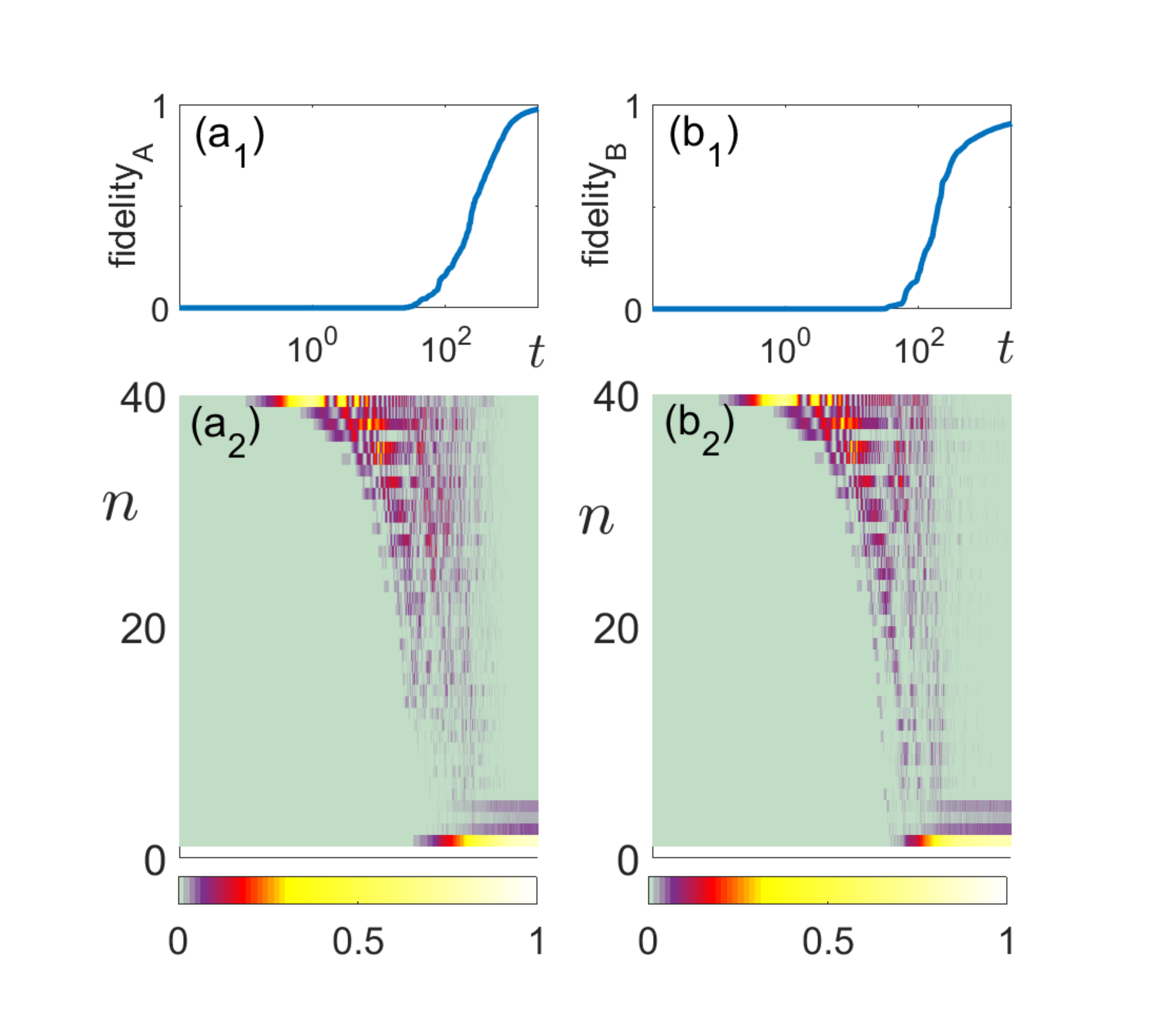}}
		\vspace{-0.5cm}
		\caption{ ($a_1$) and ($a_2$) are the fidelity and population of the excitation versus time in Lyapunov 
			control-A and ($b_1$) and  ($b_2$) are those in Lyapunov control-B. The initial excitation locates 
			at site-40, namely,  the end opposite to the manly distribution of the target state. The other 
			parameters are same to those in Fig.\ref{f:Adersonfid} (a) and (b).}
		\label{f:Adersonrobust}
	\end{figure}
\end{center}

To exhibit the control results, we use the fidelity $|\langle\psi_f|\psi(t)\rangle|^2$ between the state at 
time $t$ and the goal state to exhibit the control performance in Fig.\ref{f:Adersonfid}. When the fidelities 
approach 1, the control fields fade away synchronously in both control processes at the terminated time. But 
with the same parameters, i.e., the identical $V_c$, K, X and Z in the two control Hamiltonians, Lyapunov 
control-A (hopping-control Hamiltonian) performs faintly better than Lyapunov control-B (on-site control 
Hamiltonian). The performance difference may result from that the distribution of the goal state 
can be reached more efficiently when the hopping is modulated in the designed way. When the on-site potentials are modulated, the mainly way to reach the goal state is the hopping configuration in the free system Hamiltonian~(\ref{eq:Hamiltonian}) which is fixed. However there are other factors influence the control performance,
 for example, the form of the control Hamiltonians and the amplitudes and combination of the parameters in the 
 control Hamiltonians. Optimal control strategies may be employed to improve the control performance, for example, choosing appropriate combinations of $V_c$,  K, X and Z in the control Hamiltonians which is beyond the scope of 
 this work.

In detail, we show the evolution for the excitations on the chain during the Lyapunov control processes in 
Fig. \ref{f:Adersonrobust}. Compared to the ergodic oscillation without control when the initial excitation 
locates on one site on the chain in Fig.\ref{f:Adersondynamics} (a), the state is steered to the goal state 
in both control procedures.
\begin{center}
	\begin{figure}
		\includegraphics[scale=0.19]{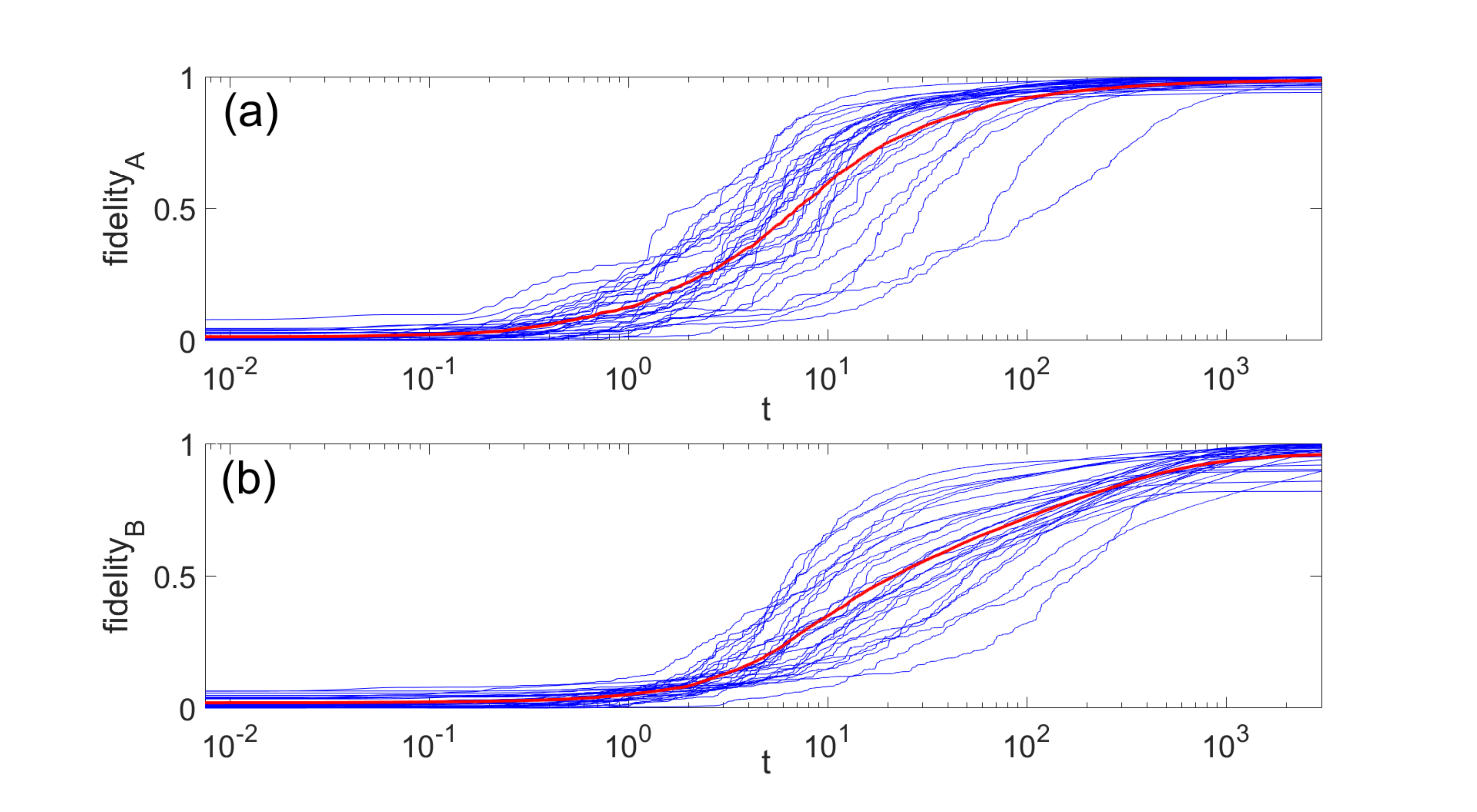}
		\vspace{0cm}
		\caption{We show the fidelity as a function of time for Lyapunov control-A and -B with 30 different 
			initial states. The amplitudes for the initial states are randomly distributed on the chain for each simulation. The red curves in (a) and (b) are average for the blue curves in each panel. The other 
			parameters are same to those
			in Fig.\ref{f:Adersonfid}. }
		\label{f:Adersoninitial}
	\end{figure}
\end{center}

Even the fidelity can reach high values in both control processes, however the oscillations of the control 
fields $f_A$ and $f_B$ in Fig.\ref{f:Adersonfid} increase the complexity in operation. Thus we may adopt optimal strategies to optimize the control fields to reduce this complexity. In Lyapunov control, the control fields are 
designed to make the state converge to the objective state.  So we redesign the control fields to optimize the
control processes. And compared to the amplitudes of control fields, the sign of them ensure the state evolves monotonously to the control target. Thus we adjust the amplitudes of the control fields but maintain the sign of 
them. Here we should confirm that the control fields are redesigned but simply change the amplitudes of those
specified by Lyapunov control method. Since the state approaches the target, the amplitudes of the control fields
become small. The alternating frequency of the sign for the control fields would be sensitive to the amplitudes 
of them. Thus we choose the envelope function with absolute value decreasing monotonically to modulate the control 
fields as bellow:
\begin{eqnarray}
\begin{aligned}
M(t)=e^{-\xi\sqrt{t}},
\end{aligned}
\label{modulation}
\end{eqnarray}
where $\xi$ is a positive constant. So the modulated control field would be
\begin{eqnarray}
f_M(t)=\left \{
\begin{array}{rl}
M(t),~~~ f_o(t)>0, \\
-M(t),~~~f_o(t)<0, \\
\end{array}
\label{bangbang}
\right.
\end{eqnarray}
where $f_M(t)$ denote the modulated control fields.  $ f_o(t)$ is the not-modulated control field designed 
by Lyapunov control method at time $t$. In this situation, the sign of $ f_o(t)$ and the modulated ones $f_M(t)$ 
are identical to make the state converge to the target monotonously. $f_{AM}$ and $f_{BM}$ will be used to denote
 the redesigned control fields corresponding to Lyapunov control-A and -B cases.  By this modulation, the control
procedure mainly becomes controlling the time interval between sequential pulses since the envelope function is 
given. This reduces the control complexity in practice.  From Fig.\ref{f:Adersonfid}, we can see that the fidelity 
can reach similar high values by the optimized control fields at the same terminated time to the not-modulated cases.

For generality, we choose 30 different initial states to check the control results in Fig.\ref{f:Adersoninitial}. 
The initial states are normalized and projecting to the sites with randomly amplitudes. It can be seen that the 
fidelity can reach high value with both control Hamiltonians  and Lyapunov control-A performs slightly better than Lyapunov control-B in average.
\begin{center}
	\begin{figure}
		\includegraphics[scale=0.25]{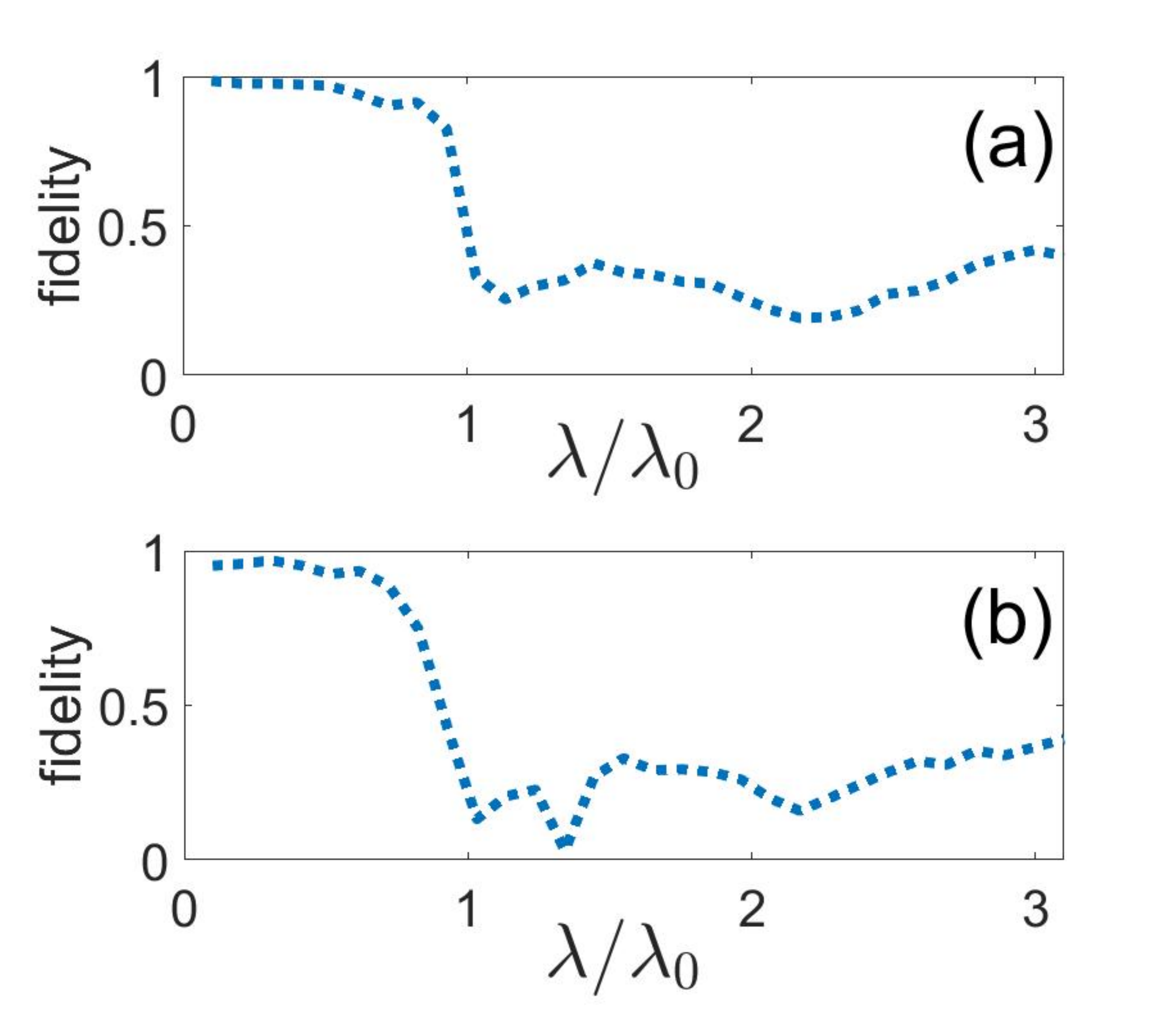}
		\vspace{0cm}
		\caption{(a): The average fidelity at time $t=2000$ versus $\lambda/\lambda_0$ under Lyapunov control-A. 
			(b): The fidelity at time $t=2000$ versus $\lambda/\lambda_0$ under Lyapunov control-B. These two 
			curves are averaged over 60 simulations with initial states whose amplitudes randomly distribute on 
			the chain. The other parameters are same to those in Fig.\ref{f:Adersonfid}.}
		\label{f:AdersonLLocal}
	\end{figure}
\end{center}

As mentioned above, the localization condition influences the transport properties of this AAH model. Since 
Lyapunov control based on transportation of the excitations, it is natural to conjecture the  Lyapunov control 
performance would be suppressed in the localized region. Thus we choose the same control parameters and initial 
state as mentioned above to examine the influence of localization. The control goal state is still the 25th edge eigenstate for the chain with $N$=40. In the parameter range $\lambda/\lambda_0\in[0.1,3.1]$, we show the average 
control results for Lyapunov control-A and -B processes at time $t$=2000 in Fig.\ref{f:AdersonLLocal}. It is obvious 
that in the localized region, namely $\lambda/\lambda_0>$1, the control effect have been suppressed in both Lyapunov control-A and -B processes. The suppression may be interpreted as that the transportation of the excitation has
been blocked in the localized region. So not only in the edge state pumping processes but also in the Lyapunov 
control processes, the localization constrain the transportation of the excitations on the chain which results to obviously different performance of the dynamical processes in nonlocal and localized regions.
\section{Discussions}
\label{DISCUSSIONS}

\subsection{Occupation imbalance and entropy as the localization indicator}\label{EntropyEO}
\begin{center}
  	\begin{figure}
  		\centering
  		{\includegraphics[scale=0.42]{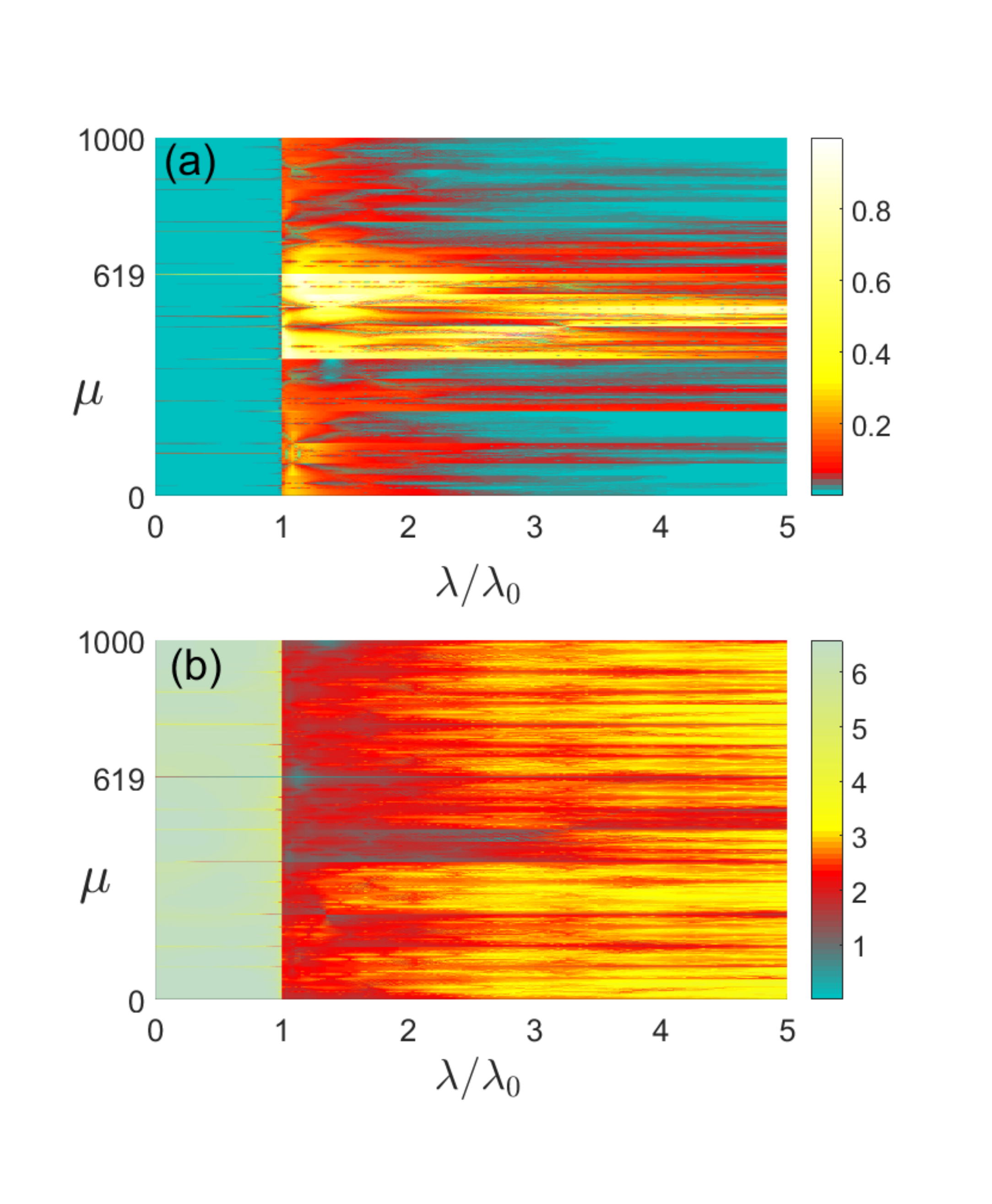}}
  		\vspace{-1cm}
  		\caption{(a): The occupation imbalance $\mathcal{I}$ for the eigenstates (denoted by $\mu$) as a function 
  			of $\lambda/\lambda_0$ for the extended AAH chain with $N=1000$.  (b): The entropy $\mathcal{E}$  for 
  			the eigenstates as a function of $\lambda/\lambda_0$ for the same chain. }
  		\label{f:AdersonEO}
  	\end{figure}
  \end{center}

Even IPR can be employed to indicate the degree of localization, it can not be a perfect indicator to show various 
kinds of localization patterns. Thus besides IPR, we also use the quantity called occupation imbalance to reveal 
the localization transition. It is defined as: $\mathcal{I}=|N_e-N_o|$, where $N_e$ and $N_o$ are the summation 
for the particle density projecting on even and odd lattice sites.  This quantity is similar to the one employed 
in experiments\cite{Science349842,NP13460} to measure ergodicity for a quantum system. Since the difference of 
occupation $N_e-N_o$ tends to vanish for even-distributed occupation pattern but finite for states tend to projecting on 
even or odd sites. Thus $\mathcal{I}$ reflects the distribution characters for the states in terms of the occupations 
on even and odd sites. Besides, the entropy defined as $\mathcal{E}=Tr[\rho\log\rho]$ is also adopted to reflect 
the distribution characters, where $\rho$ is the  density matrix and $Tr[\bullet]$ denotes the trace of $\bullet$. Considering entropy reflects the degree of confusion for the distribution of a state,  $\mathcal{E}$  would be small 
when the normalized distribution tends to flock together. Conversely, it would be large. We show $\mathcal{I}$ and $\mathcal{E}$ as a function of $\lambda/\lambda_0$ for all eigenstates of a chain with $N$=1000 in Fig.\ref{f:AdersonEO}. It can be seen 
that localization transition occur in terms of $\mathcal{I}$ and $\mathcal{E}$ which are consistent with that
in terms of IPR. A feature reflected in Fig.\ref{f:AdersonIPR} is that with increasing of $\lambda/\lambda_0$ ($v/\lambda_0$ is fixed), the degree of localization for the eigenstates tends to decrease. Correspondingly, the occupation imbalance between even and odd sites tends to decrease and the degree of confusion tends to increase with increasing of $\lambda/\lambda_0$ ($v/\lambda_0$ is fixed). These behaviors may result from that the hopping 
interaction has the effect of dispersing the localization.

\subsection{Realization of the model by cold atoms trapped in quasiperiodic optical lattice}\label{LATTICE}
Even the extended AAH model originally generates from solid physics\cite{ppslsa68874,aips3133}, however the 
necessary intense magnetic field is unavailable in such systems currently. Fortunately the developing 
quantum simulation based on cold atoms trapped in optical lattice provides platforms to study such many-body systems\cite{RMP80885,NP8267,Nature453895}. The 1D optical lattice for the extended AAH model can be realized by superimposing an optical lattice with lattice constant $\pi/k_2$ onto a primary one with incommensurate lattice
constant $\pi/k_1$. The potentials for such a superimposed lattice can be described by :
  \begin{eqnarray}
V(x)=s_1\sin^2(k_1x+\theta^{\prime})+s_2\sin^2(k_2x+\theta_v^{\prime}),
\label{Vx}
\end{eqnarray}
 where $x$ denotes position along the chain. $\theta^{\prime}$ and $\theta_v^{\prime}$ are free phases introduced 
 to describe the shift between the lattices. $s_1$ and $s_2$ ($s_1\gg s_2$) are the amplitudes in units of the 
 recoil energy $E_r=k_1^2/(2m)$, $m$ is the effective mass for atoms. In this case, the Hamiltonian reads:
\begin{eqnarray}
 H=\frac{-\nabla_x^2}{2m}+V(x).
\end{eqnarray}
After second quantization\cite{RMP80885,PRA90063638} in terms of Wannier functions, the on-site incommensurate 
modulation in Hamiltonian (\ref{eq:Hamiltonian}) is obtained when $v\sim s_2$,
 $\theta_v=\theta_v^{\prime}$ and $b=k_2/k_1$. Here we focus on the refinement for the hopping modulations to some 
 extent. We expect the hopping modulation is amended by the displacement of the extremum for the lattice potential 
 $V(x)$ in (\ref{Vx}) around the sites: $x_n=n\pi/k_1$ where $n$ denote the positions for the sites. Since tunneling
  between sites depends on distance minus-exponentially, such a relation can be assumed as: $J_{n,n+1}\sim e^{-\zeta(x_{n+1}-x_n)}$,
  where $\zeta=\frac{k_1}{2}\sqrt{\frac{s_1}{E_r}}$ when $s_1\gg E_r$ \cite{RMP80885}. Then the deviation of the 
  position for extremum of $V(x)$ in (\ref{Vx}), to the first order  in $s_2/s_1$ reads:
  \begin{eqnarray}
  \delta x_n=-\frac{s_2b\sin(2nb\pi+2\theta_v^{\prime})}{2\cos(2\theta^{\prime})s_1k_1}-\frac{\tan(2\theta^{\prime})}{2k_1\cos(2\theta^{\prime})}.
\end{eqnarray}
In this case, one gains:
 \begin{eqnarray}
x_{n+1}-x_n=\frac{\pi}{k_1}-\frac{s_2b\sin(b\pi)}{\cos(2\theta^{\prime})s_1k_1}\cos(2n\pi b+\theta_v),
\end{eqnarray}
where $\theta_v=\theta_v^{\prime}+b\pi$ is the one in Hamiltonian (\ref{eq:Hamiltonian}). Thus the hopping terms 
in the lowest order of $s_2/s_1$ are
 \begin{eqnarray}
J_{n,n+1}\sim1+\frac{s_2b\sin(b\pi)}{2\cos(2\theta^{\prime})\sqrt{s_1E_r}}\cos(2n\pi b+\theta_v).
 \end{eqnarray}
This coincides with the incommensurate modulations of hopping in Hamiltonian (\ref{eq:Hamiltonian}) when 
$\lambda\sim s_2b\sin(b\pi)/(2\cos(2\theta^{\prime})\sqrt{s_1E_r})$. Thus the extended AAH model can be implemented 
by cold atoms trapped in quasiperiodic optical lattice.
\subsection{Realization of the model by coupled optical waveguide array system}\label{kerrnonlinear}
Besides the cold atom system in optical lattice, this extended AAH model can also be simulated in coupled 
optical waveguide arrays \cite{prl103013901,prl109106402,oe146055,jpb43163001,prl102153901,prl100094101,nature424817,prl100013906,prb75205120}.
The dynamical processes is mapped to the propagation of the probe light along the waveguides. Such arrays of waveguides can be manufactured on optical bulk materials by using femtosecond laser pulses~\cite{oe146055,jpb43163001} or by 
applying high resolution, large field electron-beam lithography combine with reactive ion etching technique on AlGaAs substrate~\cite{prl103013901}.  The hopping can be tuned by varying the spacing between the waveguides which 
determines $\lambda$, $\theta$ in Hamiltonian (\ref{eq:Hamiltonian}). The on-site potentials can be modulated by 
changing the widths of the waveguides which determines $v$ and $\theta_v$ \cite{prl103013901,oe146055,jpb43163001,prl102153901}. In this apparatus, the fluorescence microscopy technique 
can be employed to observe the light intensity in the waveguides and the distribution for the wave functions
  can be examined by measuring the intensity of efferential light at the output interface.  Disorder exists 
  but small and can be factored out since the localization length associated with the disorder is much larger 
  than the spacing between the waveguides and widths of them~\cite{prl103013901}.  Since the intrinsic loss of 
  the waveguides can be turned weak and identical for all sites, it can also be factored out in this model.
\begin{center}
	\begin{figure}[htb!]
		\centering
		{\includegraphics[scale=0.28]{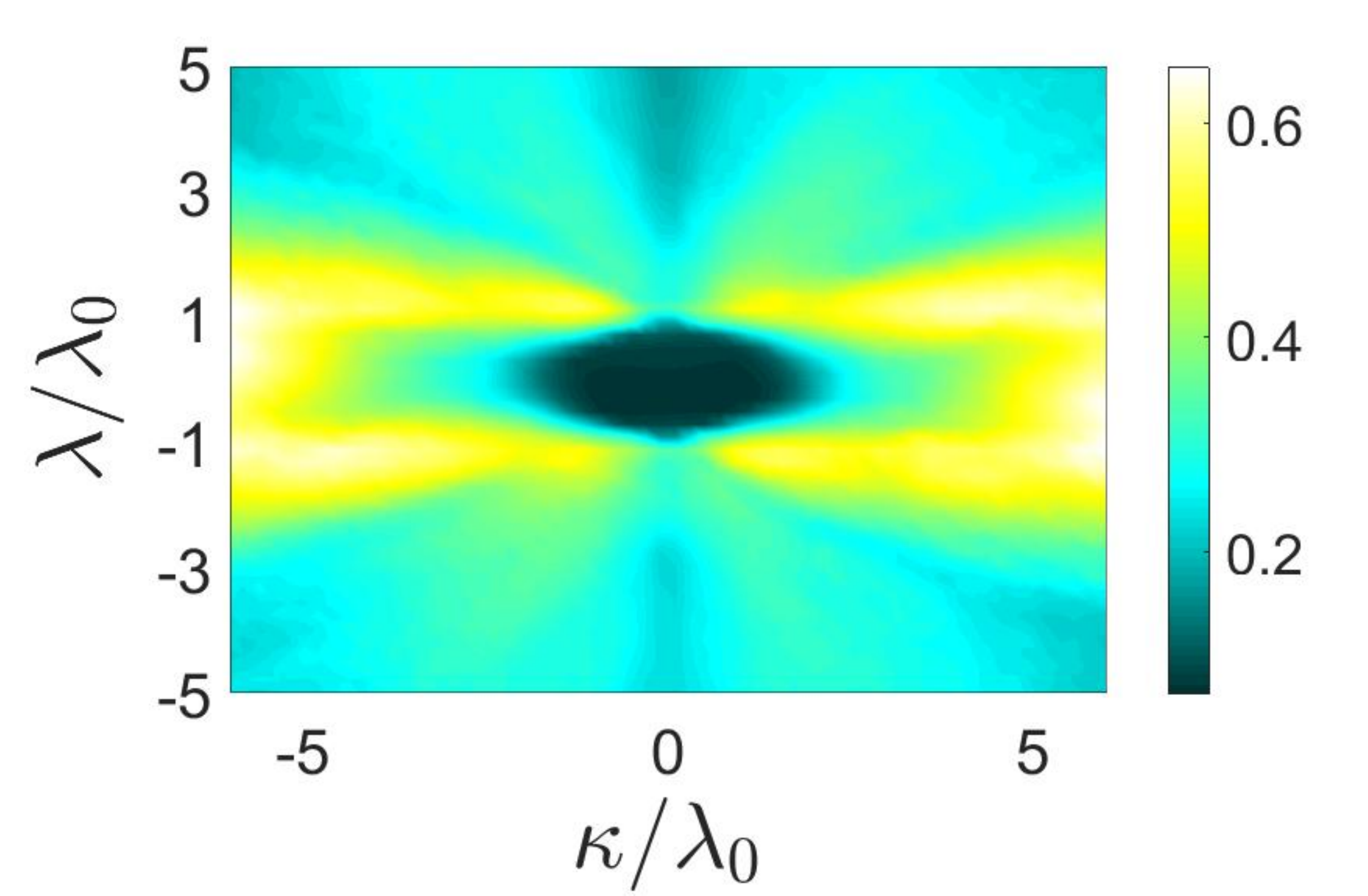}}
		\vspace{-0.5cm}
		\caption{This Figure shows
			the average IPR over $N$ simulations at time $t=100$ versus $\kappa/\lambda_0$ and $\lambda/\lambda_0$. The excitation locates on one site for each simulation.  We have set $N$=40. 
			The other parameters are same to those in Fig.~\ref{f:AdersonIPR} (b).}
		\label{f:Adersonnonlinear}
	\end{figure}
\end{center}

 The  above discussions about transport properties are in the linear scope. When one use optical waveguide 
 systems, the nonlinearity is usually considered when the intensity of the probe light injected into the system 
 is large enough. The influence of nonlinearity to some dynamical processes have been explored theoretically and experimentally \cite{prb75205120,prl100013906,prl100094101,nature424817,prl102153901,prl100013906}.
The Kerr-type nonlinearity is usually considered. The transportation for the excitation versus localization may 
change in presence of nonlinearity. The dynamical equations in presence of Kerr-type nonlinearity can be obtained 
by adding the term:
\begin{eqnarray}
\kappa|\psi_n|^2\psi_n
\end{eqnarray}
to right hand side of equations (\ref{eq:linear}) where $\kappa$ is the Kerr nonlinear coefficient indicating the 
strength of the third-order  nonlinearity ~\cite{prl100013906,prl100094101,nature424817,prl102153901,prb75205120}.
It can  be interpreted that the on-site potentials add terms proportional to $ \kappa|\psi_n|^2$ in presence of this 
kind of nonlinearity. We use the performance of transportation for the excitation on the chain for a range of $\kappa/\lambda_0$ to check the change of localization. Thus we check the average IPR at time $t=100$ over the cases 
that the initial excitations locate only on one site in each simulation in Fig.~\ref{f:Adersonnonlinear}. Namely, each site on the chain acts as a position for each initial excitation in the simulations. This is a statistical and dynamical viewpoint to check the localization property in presence of nonlinearity. And in this time interval, it is long time enough for the excitation propagating from one end of the chain to another one in the nonlocal region. From Fig.~\ref{f:Adersonnonlinear}, it can be seen that the localization condition have changed obviously in presence of nonlinearity. For example, in the original nonlocal region ($\lambda/\lambda_0\leq1$), the Kerr-type nonlinearity leads 
to localized behavior for the excitation. However, with increasing of $\kappa$, higher-order nonlinearity need to be considered. The nonlinear terms $\kappa|\psi_n|^2\psi_n$ break down to describe the dynamics~\cite{prb40546}. Since the sensitive dependence on initial conditions in nonlinear dynamical processes,  more rigorous and general influence of nonlinearity on the dynamical process maybe needed to study elusive the relation between nonlinearity and localization.

\section{ Conclusion}
\label{SUM}
In conclusion, we have studied the effect of localization transition for an extended AAH model on the adiabatic 
pumping for an edge state and  preparing the edge state by  Lyapunov control method. The on-site and hopping 
potentials are both modulated  incommensurately in this model. We show the average localization transition diagram 
over all the eigenstates as a function of system parameters. For a specified  system, not only the ground state 
but also most of the exite eigenstates exhibit localization transition. And the change of the energy spectrum 
splitting coincides with the localization transition. The influence of localization transition on dynamical processes 
are explored by two examples. Firstly, the edge state in the nonlocal region can be pumped from one end to the other 
one successfully by varying the phase parameter adiabatically. But the adiabatical pumping fails in the localized 
region. Secondly, by quantum Lyapunov control method with two different control Hamiltonians, the edge state can be prepared in the nonlocal region with high fidelities. But the control effect are suppressed in the localized region. 
These can be interpreted as that in the nonlocal region, the extended AAH chain acts as conductor for the excitation 
but insulator in the localized region.  Besides IPR, the occupation imbalance between even and odd sites and the 
entropy for the eigenstates can reveal the localization transition and coincide with IPR reveals. Finally, we propose 
that such a model can be implemented by cold atoms trapped in quasiperiodic optical lattice. Besides, the system 
composed of coupled optical waveguides can also be used  to check the discussions. In the optical waveguide system, 
we show the influence of Kerr-type nonlinearity on the localization transition in the dynamical viewpoint by means of statistical method.

\section*{ACKNOWLEDGMENTS}
We thank Prof. L. C. Wang at Dalian University of Technology for valuable discussions.
This work is supported by the National Natural Science Foundation of China (Grant No. 11534002 and 61475033).
	
\end{document}